\documentclass[twocolumn,showpacs,preprintnumbers,amsmath,amssymb,prb]{revtex4-1}

\usepackage[english]{babel}
\usepackage{graphicx}
\usepackage{dcolumn}
\usepackage{bm}
\usepackage{graphicx}
\usepackage{eepic}
\usepackage[percent]{overpic}
\usepackage[colorlinks=false]{hyperref}

\newcommand{\suml}{\sum\limits}

\newcommand{\R}{\mathbb{R}}
\newcommand{\cH}{\mathcal{H}}

\newcommand{\conj}[1]{{#1}^{\star}}
\newcommand{\deba}{\partial}
\newcommand{\hb}{\hbar}

\newcommand{\E}{E}
\newcommand{\hbtm}{\frac{\hb^{2}}{2 m_0}}

\newcommand{\ovl}[1]{\overline{#1}}

\renewcommand{\vec}[1]{\mathbf{#1}}

\newcommand{\lrsolid}{\rule[2pt]{12pt}{0.8pt}}
\newcommand{\lrdashed}{\rule[2pt]{5pt}{0.8pt}\rule[2pt]{2pt}{0pt}\rule[2pt]{5pt}{0.8pt}}
\newcommand{\lrdotted}{\rule[2pt]{1pt}{1pt}\rule[2pt]{2.5pt}{0pt}\rule[2pt]{1pt}{1pt}\rule[2pt]{2.5pt}{0pt}\rule[2pt]{1pt}{1pt}\rule[2pt]{2.5pt}{0pt}\rule[2pt]{1pt}{1pt}}
\newcommand{\lrdotdashed}{\rule[2pt]{3.5pt}{0.8pt}\rule[2pt]{2pt}{0pt}\rule[1.95pt]{1pt}{1pt}\rule[2pt]{2pt}{0pt}\rule[2pt]{3.5pt}{0.8pt}}
\newcommand{\lsdotted}{\dottedline{1.2}(0,3)(6,3)}
\newcommand{\lsdashed}{\drawline(0,3)(3,3)\drawline(5,3)(8,3)}
\newcommand{\lsdotdashed}{\drawline(0,3)(2.8,3)\drawline(3.8,3)(4.2,3)\drawline(5.2,3)(8,3)}

\graphicspath{{./figures/}}
\usepackage{siunitx,array,booktabs}
\usepackage{csvsimple}
\usepackage[para]{threeparttable}
\usepackage[dvipsnames]{xcolor}
\usepackage{colortbl}
\usepackage{etoolbox}
\robustify\tnote
\robustify\bfseries

\begin{document}

\title{The Luttinger-Kohn theory for multiband Hamiltonians: A revision of ellipticity requirements}

\author{Dmytro Sytnyk}
 \email{sytnikd@gmail.com}
 \affiliation{%
Numerical Mathematics Department.
Institute of Mathematics,
National Academy of Sciences, 
Ukraine;\\
$M^{2}NeT$ Laboratory,
Wilfrid Laurier University,
75 University Avenue West,
Waterloo, ON,
Canada, N2L 3C5.
}%

%
\author{Roderick Melnik}
 \email{rmelnik@wlu.ca}
 \affiliation{%
$M^{2}NeT$ Laboratory,
Wilfrid Laurier University,
75 University Avenue West,
Waterloo, ON,
Canada, N2L 3C5.
}%
\begin{abstract}
Modern applications require a robust and theoretically solid tool for the realistic modeling of electronic states in low dimensional nanostructures.  The $k \cdot p$ theory has fruitfully served this role for the long time since its establishment. During the last three decades several problems have been detected in connection with the application of the $k \cdot p$ approach to such nanostructures.
These problems are closely related to the violation of the ellipticity conditions for the underlying model, the fact that has been largely overlooked in the literature.
We derive ellipticity conditions for $6 \times 6$, $8\times 8$ and $14 \times 14$  Hamiltonians obtained by the application of Luttinger-Kohn theory to the bulk zinc blende (ZB) crystals, and demonstrate that the corresponding models are non-elliptic for many common crystalline materials.
With the aim to obtain the admissible (in terms of ellipticity) parameters, we further develop and justify a 
parameter rescaling procedure for $8\times 8$ Hamiltonians.
This allows us to calculate the admissible parameter sets for GaAs, AlAs, InAs, GaP, AlP, InP, GaSb, AlSb, InSb, GaN, AlN, InN. 
The newly obtained parameters are then optimized in terms of the bandstructure fit by changing the value of the inversion asymmetry parameter $B$ that is proved to be essential for ellipticity of $8\times 8$ Hamiltonian.
The consecutive analysis, performed here for all mentioned  $k \cdot p$ Hamiltonians, 
indicates the connection between the lack of ellipticity and perturbative terms describing the influence of out-of-basis bands on the structure of the Hamiltonian.
This enables us to quantify the limits of models' applicability material-wise and to suggest a possible unification of two different $14 \times 14$ models, analysed in this work.

\end{abstract}


\maketitle

\section{Introduction}
The collection of methods known as an effective mass theory is one of the fundamental topics in the physics of nanostructures. 
The theory has been used to describe a wide variety of physical phenomena ranging from the formation of electronic bands in periodic solids to the realistic field-mater interaction in modern semiconductor materials. 
%
Furthermore, the theory establishes a robust computational framework for simulating observable quantum-mechanical states and corresponding energies in the low-dimensional nanoscale systems, including quantum wells, wires and dots.

In the original Luttinger--Kohn work \cite{Lutt_55} authors applied the perturbation theory to the Schr\"odinger equation with a smooth potential and constructed a representation for valence bands Hamiltonian near the high symmetry point $\Gamma$ of the first Brillouin zone in bulk zinc blende (ZB) crystals with large fundamental bandgap.
Soon after that, Kane showed how to extend the model to the narrow gap materials  such as InSb and Ge
for instance, where one can also account for the influence of the conduction bands \cite{Kane1957}.
One of the advantages of the $k\cdot p $ theory is in its universality and flexibility when it comes to simulation of electronic transport phenomena in the presence of electromagnetic and/or thermoelastic fields\cite{Prabhakar2013,Prabhakar2012}. 
Indeed, the theory had also been extended to cover Wurtzite (WZ) type of crystals, materials with inclusions, heterostructure materials and superlatices \cite{Xia91,Prabhakar2013a,Alvaro2013}.
Another advantage of the effective mass theory is its flexibility, as one can easily adjust the models to include additional effects like strain\cite{Patil2009}, piezoelectricity, magnetic field, and respective nonlinear effects.
These inbuilt multiscale effects are  crucial for such applications as light-emission diodes, lasers, high precision sensors, photo-galvanic elements, hybrid bio-nanodevices, and many others \cite{Abajo2007}.

For a wide range of applications the Luttinger-Kohn  models have provided good, computationally feasible and
efficient approximations that agree well with experimental results \cite{ChuangB_95, ChuangB_09}.
However, for some types of crystal materials band structure calculations based on such multiband models lead
to the solutions with unphysical properties\cite{Smith_sp_86, Szmulowicz_96} or so called spurious solutions \cite{foreman_sp_07, Yang_sp_05, Veprek07, Veprek08, Melnik2009, kpEllptArxSytnyk2010}.

As a result, there have been various attempts to explain the origin of the spurious solutions and
develop some reliable procedures on how to avoid them\cite{Foreman_sp_97, Cartoixa2003, Veprek07, kpEllptArxSytnyk2010}.
 These approaches rely on three main ideas: (a) to modify the original Hamiltonian and remove the terms
 responsible for the spurious solutions\cite{Kolokolov_sp_03, Yang_sp_05}, (b) to
 change band-structure parameters\cite{Eppenga_sp_87, Foreman_sp_97, foreman_sp_07}, and (c) to identify and exclude from simulations the physically inadequate observable states \cite{Kisin_sp98, ChuangB_09} or change the numerical scheme to avoid such states altogether\cite{Cartoixa2003, Eissfeller2011}.
All mentioned approaches suffer from the common weakness -- the lack  of clear justification of the underlying theoretical procedure and thus from limitations in their applicability \cite{Veprek07, Veprek08}.

In this work we show that spurious solutions are just a consequence of a more fundamental problem in applications of the effective mass theory: the non-ellipticity of the multiband Hamiltonian in the position representation. 

The systematic study of connection between the structure of $6 \times 6$, $8\times 8$ and $14 \times 14$ Hamiltonians, their ellipticity in the position representation and the material parameters for ZB crystals allow us to conclude that the widely adopted $k \cdot p$ models turn out to be non-elliptic (hyperbolic) for a broad class of known material parameters.
The phase space of the hyperbolic model is wider than the spaces of norm-bounded observable states. 
Such models, therefore, are susceptible to unphysical solutions,  even in the bulk case. 
Meanwhile, the corersponding time-dependent Schr\"odinger equation loses the fundamental property of state conservation \cite{Landau_qm1982}.

These facts lead to an important assertion. 
Since any qualitative multiband approximation of Schr\"odinger Hamiltonian must preserve its core physical properties, such as ellipticity and, as a consequence, semi-boundedness of set of energy states;   the lack of ellipticity for certain materials implies that the usage of multiband Hamiltonian for such materials is fundamentally incorrect. 
This results in substantial ramifications for the applications of effective-mass theory to bulk solids and heterostructures. 

The whole procedure of obtaining the materials parameters from experiment and their incorporation into mathematical models of effective mass theory needs to be revisited, taking into account the general ellipticity constraints derived in the present work.
Before this is done, we propose here the sets of elliptic Hamiltonian parameters for  GaAs, AlAs, InAs, GaP, AlP, InP, GaSb, AlSb, InSb, GaN, AlN, InN, optimized in terms of the bandstructure fit. 
We also supply a parameter rescaling procedure used to obtain these sets from the available non-elliptic parameters.


The paper is organized as follows.
First, we revise basic properties of the Schr\"odinger equation and its approximations represented by $k\cdot p$ models.
In section \ref{sec6x6} we outline a mathematical procedure to obtain the ellipticity constrains for a Hamiltonian in the position representation. 
For the $k \cdot p$  Hamiltonians the constraints are comprised of the set of linear material-dependent inequalities\cite{Veprek09}.
In sections \ref{sec6x6} and  \ref{sec8x8} we present a direct evaluation of ellipticity constraints for $6\times 6$ and $8\times 8$ ZB Hamiltonians based on parameter sets gathered from major material-data sources\cite{kp7Lawaetz71, LB1, Vurgaftman2001, Madelung2004, Rinke2008}. 
Most of the 53 analyzed parameter sets lead to the failure of the Hamiltonians' ellipticity. 
That is why the main part of section \ref{sec8x8} is devoted to a parameter rescaling procedure aimed at correcting the Hamiltonian's  ellipticity.
As a result of the procedure we propose elliptic parameter sets for all analyzed materials. 
The newly obtained sets are, then, compared to the original materials parameters by means of the differences in the associated bandstructures of $8\times 8$ Hamiltonian.
In this section, we further extend the ellipticity conditions of $8\times 8$ model \cite{kp8ProperKane_Bahder90} to the case of nonzero inversion-asymmetry parameter $B$.
Afterwards, $B$ is used to improve the bandstructure fit of the proposed elliptic parameter set for indium nitride.
Section \ref{sec14x14} is devoted to the ellipticity analysis of two existing $14 \times 14$ models\cite{kp14Pfeffer96,kp8x84x46x6_sp_bc_Winkler2003}.

The summary of results together with discussions on applicability and future directions are given in the concluding section.  

\section{Overview of Luttinger-Kohn bandstructure theory }
The material properties (such as fundamental band-gaps and spin-orbit splitting energies) obtained experimentally, represent real quantum phenomena, whereas models based on multiband Hamiltonians are meant to approximate them.
As such these models are derived from the stationary Schr\"odinger equation that represents  an averaged charge carrier interactions in the crystalline structure\cite{Kane1982,VoonWillatzen2009}.
The derivation scheme involves the application of Bloch wave representation and the projection of the original Hamiltonian to the orthogonal subspace of the reduced phase space\cite{Lutt_55,BP1}. 
The projective part of Hamiltonian is then adjusted with help of perturbation theory \cite{Lutt_55, Lutt_56, BP1} to account for the influence of outer bands. 
However, this last step lacks a rigorous theoretical foundation as it does not guarantee the convergence of the perturbative expansion\cite{Kato1,MThSytnyk2010}.
The result is that the derived Hamiltonian, although directly based on the experimental parameters ( Tables \ref{tabMPZBL6x6}, \ref{tabMPZBL8x8}, \ref{tabMPZBL14x14resc}), represents a totally different mathematical object compared to its origin.
The physical evidence, to support this claim has been already known for GaAs \cite{kp14Pfeffer90}  and for Si \cite{Dorozhkin_08}.


We start with the Schr\"odinger equation 
\begin{equation}\label{SchrodEqs}
H_0\psi(x) \equiv \left (\frac{\mathbf{p}^2}{2m_0}+V(x)+H_{SO}\right )\psi(x)=E_n\psi(x),
\end{equation}
where $\mathbf{p}=i\hbar \nabla$ is a momentum operator of charge carrier with the mass $m_0$, $V(x)$ is the effective potential, $x \in \Omega \subset {\R}^3$. The unknown $E_n$, stands for the eigenenergy of the system and the function $\psi(x)$ is the corresponding eigenstate. 
The Hamiltonian $H_{SO} $ accounts for relativistic effects of spin. 

In the finite domain $\Omega$ we supplement \eqref{SchrodEqs} by the boundary conditions 
\begin{equation}\label{SEBC}
\psi(x) = f(x), \quad x \in \partial \Omega,
\end{equation}
assuming that the combination of given $\Omega$ and $f(x)$ endows operator $H_0$ with all necessary properties, postulated by the standard axiomatic approach to quantum mechanics\cite{Ballentine2006}.  
The operator $H_0$ is an elliptic partial differential operator. 
It is symmetric over its domain of definition $D(H_0) \subset \cH^3(\Omega)$. 
Furthermore we require that the boundary $\partial \Omega$ is sufficiently smooth, so that a self-adjoint extension of $H_0$ exists and possesses the property of the probability current conservation\cite{Dirac81,Ballentine2006}.
All mentioned assumptions can be satisfied in the bulk case\cite{Teschl2009}, which will be our main focus throughout the work. 

%
If $V(x)$ is a gently varying function over the unit cell\cite{Lutt_55}, the original operator $H_0$ can be approximated by another operator $H$ (using Bloch theorem), determined by the projection $P$ of $H_0$ on the considered eigenspace and L\"owdin perturbation theory \cite{Lutt_55, BP1}. 
The last step in this approximation procedure accounts for the influence of the elements from the
space (so-called class of states B) complement to the chosen eigenspace (so-called class of states A) by the formula
\begin{equation}\label{eq_PO}
  H = P H_0 + \suml_{i=1}^r \delta^r H^{(r)}
\end{equation}
up to the order $r$. Setting $\delta=1$ leads one to the final approximation, under the assumption that the series (\ref{eq_PO})
is convergent for such $\delta$.
Despite wide applicability of such approximations,
the intrinsic ellipticity requirements for the realizations of $H$ have not been explicitly verified
in a systematic manner
(see \onlinecite{Lutt_55,Kane1957,BP1,Foreman_sp_97,Burt_99}, as well as more recent works\cite{kp14Pfeffer96,kp8x84x46x6_sp_bc_Winkler2003,Kolokolov_sp_03,Yang_sp_05,foreman_sp_07,Melnik2009,Eissfeller2011}). 
The only known to us work where it has been done for the case of  InAs, GaAs and Al$_{0.3}$Ga$_{0.7}$As is [\onlinecite{Veprek07}].
Hence, in what follows we analyze such requirements systematically for all common $6 \times 6$, $8 \times 8$ and $14 \times 14$ ZB Hamiltonians.

\section{Six-bands Hamiltonian analysis}\label{sec6x6}
This section is devoted the ellipticity analysis of the classical $6 \times 6$ Hamiltonian for ZB \cite{Lutt_55} type of crystals, demonstrating our approach in detail.
In this work we use the Luttinger parameter notation which is common in recent works on the subject. When necessary, the parameters will be converted from other parameter notations\cite{VoonWillatzen2009}

The Luttinger-Kohn (LK) Hamiltonian is defined as follows\cite{Lutt_56,BP1}
\begin{widetext}
\begin{equation*}
  \label{eqZBBPHam}
  H{^{LK}}=\left(
  \begin{array}{ccccccc}
 P+Q & S & R & 0 & -\frac{1}{\sqrt{2}} S & -\sqrt{2} R\\
 \conj{S} & P-Q & 0 & R & \sqrt{2} Q & \sqrt{\frac{3}{2}} S\\
 \conj{R} & 0 & P-Q & -S & \sqrt{\frac{3}{2}} \conj{S} & -\sqrt{2} Q\\
 0 & \conj{R} & -\conj{S} & P+Q& \sqrt{2} \conj{R}& -\frac{1}{\sqrt{2}}\conj{S} \\
 -\frac{1}{\sqrt{2}} \conj{S} & \sqrt{2} Q & \sqrt{\frac{3}{2}} S & \sqrt{2}R & P- \Delta_{SO} & 0\\
 -\sqrt{2} \conj{R} & \sqrt{\frac{3}{2}} \conj{S} & \sqrt{2}Q
 & -\frac{1}{\sqrt{2}} S & 0 & P- \Delta_{SO}\\
  \end{array}
    \right)
    \quad
    \begin{array}{l}
  P=-\frac{\hb^{2}}{2 {m}_{ {0}}} \gamma_1 \mathbf{k}^2,\\[11pt]
  Q=-\frac{\hb^{2}}{2 {m}_{ {0}}} \gamma_2(k_x^2+k_y^2 - k_z^2),\\[11pt]
  R=-\frac{\hb^{2}}{2 {m}_{ {0}}} \frac{-\sqrt{3}}{2}\left[(\gamma_2+ \gamma_3)k_{-}^2+(\gamma_2- \gamma_3)k_{+}^2\right],\\[11pt]
  S=-\frac{\hb^{2}}{2 {m}_{ {0}}}(-2\sqrt{3})\gamma_3 k_{-} k_z,
  \end{array}
\end{equation*}
\end{widetext}
where $\vec{k}^2 = k_x^2 +  k_y^2 + k_z^2$, $k_{\pm}=k_x\pm i k_y$.
Each of the $P, Q, R, S$ is a second order position dependent differential operator in the position representation or, equivalently, second order polynomial in the momentum representation \cite{Lutt_55}.

Our aim is to check the type (elliptic, hyperbolic or essentially hyperbolic) of the $H^{LK}$ as a
partial-differential operator (PDO), keeping in mind that the Schr\"odinger operator
from (\ref{SchrodEqs}) is elliptic.
Only the second order derivative terms are playing the dominant role in the following analysis because
contributions from the terms linear in the components of $\mathbf{k}$ as well as from the potential, are bounded in the domain $D(H^{LK})$ \cite{Hormander2}.
It means that the results for more complicated physical models with potential contributions from
additional fields (e.g. strain, magnetic field, etc.) will stay the same as for the original $H^{LK}$, analyzed here.
The fact that the Hamiltonian is a linear operator guarantees that it is also true for any
other representation of $H^{LK}$ obtained by linear (basis) transformations.

In a more general sense, for any $m$--dimensional matrix PDO $H=\{h_{ij}\}_{i,j =1}^{m}$,
where each element $h_{ij}$ is a second order one--dimensional PDO \cite{Egorov11998, Hormander1}
\begin{equation}\label{pdo}
h_{ij} = \suml_{k,l =0}^n h_{ij}^{kl} \frac{\deba^2}{\deba x_k \deba x_l},
\end{equation}
the associated quadratic form (also known in the mathematical literature as a principal symbol ) is defined by
\begin{equation}\label{QF}
G(\xi_1, ..., \xi_{nm}) = v M v^T, \quad v = \left(\xi_1, \ldots , \xi_{nm}\right),
\end{equation}
where $M$ is an $mn \times mn$ matrix composed from the elements $h_{ij}^{kl}$.
The $\mathrm{k\cdot p}$ Hamiltonians in $\R^n$ are a special case of (\ref{pdo}). 
They are symmetric as a matrix PDO so the associated quadratic form $G$ will have $M$ with only real eigenvalues $\lambda_i$
(e. g. \onlinecite{Veprek07}).

Using these notations, the procedure of obtaining the ellipticity condition for $H$ is reduced to the question about
the sign of $\lambda_i$ for the associated $M$.
More precisely, the matrix differential operator $H$ will be elliptic if and only if all eigenvalues
of the corresponding Hermitian $M$ will have the same sign \cite{Hormander2, Egorov11998}.

In general, it is a challenging task to calculate the eigenvalues of $M$ explicitly, even for Hamiltonians with dimension as small as $3\times 3$, but this has proved to be possible\cite{MThSytnyk2010} for highly symmetric and sparse band structure Hamiltonians like $H^{LK}$ and several others considered here.

Taking into account the fact that the sequence of eigenenergies of $H_0$ is semi-bounded from below, for an approximation $H^{LK}$ we obtain
\begin{equation}\label{eqineig}
\lambda_i<0, \quad \ i=0,1,\ldots, nm.
\end{equation}
Constrains \eqref{eqineig} guarantee the ellipticity (in strong sense  \cite{Hormander1}) of Hamiltonian $H$.
The operator $H$ possesses a self-adjoint extension in $D(H)\subset \cH^{n+2}(\Omega)$, $n>0$,
%
provided that the boundary $\partial \Omega$ is sufficiently smooth, as we have assumed in the previous section. 
Then it can be extended to a Hermitian operator by a closure in the norm [p. 113, \onlinecite{Egorov11998}] or via the Lax-Milgram procedure \cite{Hormander2}.
From the physical point of view the smoothness characteristics of $D(H)$ fulfil the natural assumption of quantum theory that the state of the system must be
a continuous function of spatial variables even when some coefficients of $H$ have finite jumps,
as it is the case for heterostructures consisting of different materials \cite{ChuangB_95, Burt_99}.
The direct calculation by (\ref{QF}) for $H^{LK}$ ($n=3$, $m=6$) leads us to the $18 \times 18$ matrix $M^{LK}$
with the following distinct eigenvalues:
\begin{equation}\label{eig_LK}
\begin{array}{cc}
\hspace*{-8pt}
{\lambda_1}=-{E}(\gamma_1 + 4\gamma_2 + 6\gamma_3),&  
{\lambda_{2}} = {E}(3\gamma_3 -\gamma_1 - 4\gamma_2),\\[2pt]
\hspace*{-15pt}
{\lambda_{3}} = {E}(2\gamma_2-\gamma_1  + 3\gamma_3),& 
{\lambda_{4}} = {E}(2\gamma_2 	-\gamma_1  - 3\gamma_3),
\end{array}
\end{equation}
where $\lambda_1$, $\lambda_{2}$, $\lambda_{3}$, $\lambda_{4}$ have the multiplicity 2, 4, 6 and 6, respectively,  $\E = \hbtm$ and $\gamma_1, \gamma_2, \gamma_3$ are the Luttinger material parameters mentioned above. 
By substituting (\ref{eig_LK}) into (\ref{eqineig}), we receive the system of linear  inequalities with respect to  $\gamma_1$, $\gamma_2$, $\gamma_3$.
They describe the feasibility region $\Lambda_{-}$ in the space of ordered triplets $\gamma_1, \gamma_2, \gamma_3$. 
In this work we shall call a triplet of numbers $a, b, c$ feasible if  $(a, b, c) \in \Lambda_{-}$. 
More generaly, we call a set of material parameters admissible if the Hamiltonian based on this set is an elliptic  partial differential operator. 

\begin{figure}[ht]
  \begin{center}
        \includegraphics[width=0.9\linewidth]{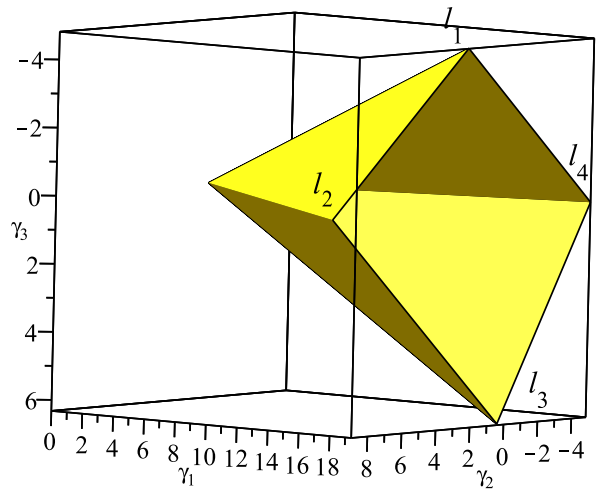}
  \end{center}
  \caption{The part of the boundary of the feasibility region $\Lambda_{-}$ along with the edges $l_1,l_2,l_3,l_4$ (color online).}
  \label{fig:DLK}
\end{figure}

When $(\gamma_1, \gamma_2, \gamma_3) \in \Lambda_{-}$ the Hamiltonian $H^{LK}$ is an elliptic PDO with the semi-bounded sequence of eigenvalues.
One can use similar reasoning to obtain the corresponding inequalities for other common representations of $H^{LK}$ like those through the parameters $A, B, C$ \cite{Lutt_55}. Evidently, any solution of (\ref{eqineig}) for (\ref{eig_LK}) would have a unique corresponding solution in the $A, B, C$ notation\cite{kpEllptArxSytnyk2010} (the aforementioned ellipticity analysis in full detail is presented in [\onlinecite{MThSytnyk2010}]).
The region $\Lambda_{-}$ comprises an unbounded pyramid in $\R^3$ (cf. Fig. \ref{fig:DLK}) with
the following rays as its edges:
$$  \begin{array}{ll}
    l_1 = (8 t, t , -2 t), &  l_2 = (2 t, t ,0), \\
    l_3 = (3 t, 0, t), &  l_4 = (4 t, -t , 0),
  \end{array}
$$
where $t \in [0, \infty)$, and the vertex is situated at the origin $\gamma_1=\gamma_2=\gamma_3=0$.
The boundary of $\Lambda_{-}$ and the edges $l_1,l_2,l_3,l_4$ are illustrated in FIG. \ref{fig:DLK}.

%
\sisetup{
	round-mode      = places,
	round-precision = 3,
	table-number-alignment = left,
	table-figures-integer  = 2,
	table-figures-decimal  = 3,
	table-align-text-post = false,
}

To determine ellipticity of $H^{LK}$, we gathered in Table \ref{tabMPZBL6x6} the material parameters $\gamma_1, \gamma_2, \gamma_3$ for GaAs, AlAs, InAs, GaP, AlP, InP, GaSb, AlSb, InSb, GaN, AlN, InN, C and evaluated  $\lambda_1$, $\lambda_{2}$, $\lambda_{3}$, $\lambda_{4}$ for the gathered triplets.
As it turned out, the eigenvalues $\lambda_1$, $\lambda_{2}$, $\lambda_{4}$ are negative for all analysed parameter sets.
In that case the ellipticity is determined by the value $\lambda_{3}$.
We provide the values of $\lambda_{3}/E$ along with two other parameter dependent quantities which are important for the current work's ellipticity analysis.
The first is the distance $d$ from the parameter triplet $(\gamma_1, \gamma_2, \gamma_3)$ to $\Lambda_{-}$.  
Second is an absolute ratio $\rho$ between positive and negative values of $\lambda_1$, $\lambda_{2}$, $\lambda_{3}$, $\lambda_{4}$. 
\begin{table*}
	\caption{The material parameters for ZB type crystals, $d$ -- distance from
		the point $(\gamma_{1},\gamma_{2},\gamma_{3})$ to the feasibility region $\Lambda_{-}$}
	\label{tabMPZBL6x6} \tabcolsep=0.4em 
	\begin{threeparttable}
		\csvloop
		{
			file=ZB_6x6_gamma_lambda_d_rhoSS.csv,
			no head, column names reset, column count=25,
			column names={%
				2=\one, 3=\two, 4=\three, 5=\four, 6=\five, 7=\six, 8=\seven, 9=\eight ,10=\nine, 11=\ten, 12=\eleven, 13=\twelve, %
				13=\thirteen, 14=\fourteen, 15=\fifteen, 16=\sixteen, 17=\seventeen, 18 = \eighteen, 19 =\nineteen, 20=\twenty, 21=\twentyone, 22=\twentytwo, 23=\twentythree, 24=\twentyfour},
			before reading=\centering\sisetup{table-number-alignment=center},
			tabular={r|l|SSS|SS[table-figures-integer = 1, table-figures-decimal = 3, round-precision = 3]S[table-figures-integer = 1, table-figures-decimal = 3, round-precision = 3]|r|l|SSS|SS[table-figures-integer = 1, table-figures-decimal = 3, round-precision = 3]S[table-figures-integer = 1, table-figures-decimal = 3, round-precision = 3]},
			table head=\toprule[0.3pt]\midrule[0.3pt]\bottomrule %
			\# & El %
			& {$\gamma_1$} & {$\gamma_2$} & {$\gamma_3$} %
			& {$\lambda_3/E$} & {$d$} & {$\rho$} %
			& \# & El %
			& {$\gamma_1$} & {$\gamma_2$} & {$\gamma_3$} %
			& {$\lambda_3/E$} & {$d$} & {$\rho$} %
			\\\hline\midrule\bottomrule, 
			command=%
			\thecsvrow & \one\text{\eleven} %
			& \two & \three & \four %
			& \seven & \nine & \ten%
			& \the\numexpr\value{csvrow}+22\relax & \fourteen\text{\twentyfour} %
			& \fifteen & \sixteen & \seventeen %
			& \twenty & \twentytwo & \twentythree,
			filter = \(\thecsvinputline>1\) \and \(\thecsvinputline<25\),
			table foot=\toprule\midrule[0.3pt]\bottomrule[0.3pt],
		}
		\begin{tablenotes}\footnotesize
			
			\item [a] Set 1 from \onlinecite{LB1}
			\item [b] Set 2 from \onlinecite{LB1}
			\item [c] Set 3 from \onlinecite{LB1}
			\item [d] Set 4 from \onlinecite{LB1}
			\item [e] Set 5 from \onlinecite{LB1}
			\item [f] Set 6 from \onlinecite{LB1}
			\item [g] Obtained by extrapolation from 5-level model
			\item [h] Measured at $T = 300K$
			\item [i] Set from \onlinecite{Madelung2004}
			\item [k] The sets from \onlinecite{Vurgaftman2001} \onlinecite{LB1}
		\end{tablenotes}
	\end{threeparttable}
\end{table*}

From Table \ref{tabMPZBL6x6} one can observe that among all analysed materials only carbon 
has admissible sets of parameters ( the last two sets from of Table \ref{tabMPZBL6x6} indicated by 0 in the $\rho$ column ).
All other gathered parameters yield $\lambda_3>0$. That is why the Hamiltonian $H^{LK}$ is not elliptic for the corresponding materials. It may even have no symmetric domain $D(H^{LK})$ as opposed to the original partial differential operator $H_0$. 
Moreover, instead of the inclusion $D(H^{LK})=D(H)\subset D(H_0) \subset \cH^{3}(\Omega)$ we have only
\begin{equation}\label{DLKnonel}
D(H^{LK})=D(H)\subset \cH^{1}(\Omega).
\end{equation}
It means that the discontinuous solutions of (\ref{SchrodEqs}) are theoretically possible. 
They will occur in the models with jump discontinuous coefficients\cite{Courant1956}, which is the case for heterostructure materials.
Additionally, the double degeneracy of $\lambda_3>0$ from (\ref{eig_LK}) means that for certain $\Omega$ there exists a two-dimensional manifold of $D(H^{LK})$ with non-physical in terms of \eqref{DLKnonel} solutions to \eqref{SchrodEqs}. 
Thus, the momentum operator from (\ref{SchrodEqs}), will be ill-defined for such eigenstates of $H^{LK}$ (by the embedding theorems, [p. 119, \onlinecite{Egorov11998}]).
All the above arguments allow us to conclude that the $H^{LK}$ does not provide a sufficiently good approximation to $H_0$, preserving the type of the PDO, for the most of available data.

Let us return to the feasible parameters from Table \ref{tabMPZBL6x6}. 
For carbon the parameter values were analyzed earlier\cite{Reggiani1983} and it was noted that they don't agree well with the Hall effect experimental measurements. 
In the earlier work\cite{kpEllptArxSytnyk2010} we showed that experimentally consistent sets for C are not admissible in terms of ellipticity. 

Concerning the rest of the materials from Table \ref{tabMPZBL6x6} we observe a clear correlation between the average distance to $\Lambda_{-}$ per material and the size of the fundamental bandgap.
Namely the sets for the large-bandgap materials: AlP, AlAs, GaP, GaN, InP  are noticeably close ($d<1$) to $\Lambda_{-}$.
The closest in terms of the distance set number 19 for AlP can be made elliptic by the direct adjustment.
Other materials have smaller gap and as a consequence are further away.
The average distance to $\Lambda_{-}$ for GaAs is around 1.7. For InAs the distance is more than 6. 
The indium antimonide is an extreme case here, having distance of more than 12. 
This material has the smallest bandgap and the high curvature of light-hole bands.  
It is known from the experiments\cite{Kane1957,Pidgeon1966,Cardona1986} that the valence-band-only Luttinger-Kohn  model is insufficient for InSb like materials, and presented analysis support this fact theoretically. 
The ellipticity of the higher band $k \cdot p$ models are considered in the next sections.

\section{Eight-band Hamiltonians}\label{sec8x8}
This section is devoted to the analysis of Kane model\cite{Kane1982,kp8ProperKane_Bahder90}.
The basis set of $8 \times 8$ Kane Hamiltonian\cite{kp8ProperKane_Bahder90} contains two more elements $\left|S\uparrow\right\rangle$ and  $\left|S\downarrow\right\rangle$ in addition to the basis set of $H^{LK}$.
These new elements of basis  represent the influence of the innermost conduction band. 
Recall that the influence of the out-of-basis states is again treated perturbatively up to the second order by using the L\"owding perturbation theory.  
In this section we will follow the exposition of [\onlinecite{kp8ProperKane_Bahder90}], because it presents the most general description of $8\times8$ Kane Hamiltonian for zinc blende crystals.
Naturally, the results presented here remain valid\footnote{\label{article_codes} All computations were performed in Maple. Codes are available at \url{www.imath.kiev.ua/~sytnik/projects/kp}} for other  versions\cite{VoonWillatzen2009,Birner2014,Bastos2016} of the same Hamiltonian.   
Since our main focus is to check the ellipticity conditions we shall drop the spin-orbit interaction part, labelled as $H_{s.o.}+H'_{s.o.}$  in eq. (13) from [\onlinecite{kp8ProperKane_Bahder90}]. This part of the  Hamiltonian is linear in $k$ and therefore won't affect the form of $G$.  
(as we have mentioned before, only second order terms in $k$ are essential for ellipticity analysis) 
Then, following Kane\cite{Kane1982}, we rewrite the resulting operator in the block-diagonal form 
$$
H^K= \left(
      \begin{array}{cc}
        H^K_\uparrow & 0 \\
        0 & H^K_\downarrow \\
      \end{array}
    \right),
$$
where $H^K_\uparrow$ is the Kane $4\times4$ interaction matrix \cite{Kane1957}, given by \eqref{eqHKane4} in the basis $\left|S \uparrow \right\rangle, \left|X\uparrow\right\rangle, \left|Y\uparrow\right\rangle, \left|Z\uparrow\right\rangle$\cite{kp8ProperKane_Bahder90}. 
The matrix $H^K_\downarrow$ that is also defined by \eqref{eqHKane4}, acts upon the spin-down part of the basis $\left|S \downarrow \right\rangle, \left|X\downarrow\right\rangle, \left|Y\downarrow\right\rangle, \left|Z\downarrow\right\rangle$.
\begin{widetext}
\begin{equation}\label{eqHKane4}
    H^K_\uparrow = \left(
      \begin{array}{cccc}
        E_c +\left (E+A'\right )\vec{k}^{2}&
        iP_0 k_x +Bk_{y}k_{z}&
        iP_0 k_y +Bk_{x}k_{z} & 
        iP_0 k_z +Bk_{x}k_{y}\\[12pt]
        -iP_0 k_x+Bk_{y}k_{z} & 
        \parbox{2.5cm}{$E_v+ M' (k_y^2 + k_z^2)$ $+L' k_x^2 +\E \vec{k}^2 $}& 
        N' k_x k_y & N' k_x k_z \\[12pt]
        -iP_0 k_y+Bk_{x}k_{z} & 
        N' k_x k_y & 
        \parbox{2.5cm}{$E_v +M' (k_y^2 + k_z^2)$ $+L' k_x^2  +\E \vec{k}^2 $}& 
        N' k_y k_z \\[12pt]
        -iP_0 k_z+Bk_{x}k_{y} & 
        N' k_x k_z & 
        N' k_y k_z & 
        \parbox{2.5cm}{$E_v  + M' (k_y^2 + k_z^2)$\\ $+ L' k_x^2+\E \vec{k}^2 $}\\
      \end{array}
    \right).
\end{equation}
\end{widetext} 
Parameters $A',B, P_0, M', N', L'$ are known as Kane parameters\cite{Kane1982}, their definitions are provided in Table 4.2 of \onlinecite{VoonWillatzen2009}. 
The quantities $E_c$ and $E_v$ are the conduction- and  valence-band energies correspondingly, $\E$ is equal to $\hbtm$, as before. 
The parameter $A'$ represents the influence of the higher bands on the conduction band included into the basis.
The parameter $P_0$ accounts for a mixing of conduction and valence band states away from $\bf k =0$. 
$B$ is a so-called inversion asymmetry parameter. 
It is equal to  zero in the materials with centrosymmetric crystal structure like diamond \cite{kp8ProperKane_Bahder90}. 
By setting $B=0$ in \eqref{eqHKane4} we obtain a simplified version of \eqref{eqHKane4} that is known as Bir-Pikus $4\times4$ Hamiltonian.
The general case of $H_K$ when $B \neq 0$ was studied by T. Bahder ( Eq. (15) in [\onlinecite{kp8ProperKane_Bahder90}]).
In practice the mentioned parameters are fitted to experimental data;
It is frequently assumed in the literature that the simplified version of $H_K$ 
provides a sufficiently good description of the physical phenomena in ZB crystals with face-centered lattice too. 
As we will later demonstrate, the Hamiltonian of such simplified model is non-elliptic for all studied material parameter sets and therefore is prone to the appearance of spurious solutions.
The parameter $B$ can not be set to zero for the materials where $E+A'<0$. 


Similarly to the $6\times 6$ case, it is common to rewrite Hamiltonian $H^K$ in the basis where its spin-orbit interaction part becomes diagonal. 
Usually one additionally pre-multiplies the original basis functions to make inter-band matrix elements and possibly other physically relevant quantities real-valued. 
%
%
%

Direct calculation of eigenvalues for the quadratic form associated with $H^K_\uparrow$,  described in details for the Luttinger--Kohn case from the previous section, gives us five distinct eigenvalues
\begin{equation}\label{eq:eigQFKP4x4}
\begin{array}{rl}
\lambda'_1 &= \E + L'+N', \\ 
\lambda'_2 &= \E + L'-\frac{1}{2}N', \\ 
\lambda'_3 &= \E + M'-\frac{1}{2}N', \\[2pt]
\lambda'_4 &= \E + \frac{2 A' + 2M'+N'}{4} - \sqrt {\frac{(2 A'- 2M'-N')^2}{16} + \frac{B^2}{2}},\\[2pt]
\lambda'_5 &= \E + \frac{2 A' + 2M'+N'}{4} + \sqrt {\frac{(2 A' - 2M'-N')^2}{16} + \frac{B^2}{2}}.\\[2pt]
\end{array}
\end{equation}
The presence of the second order conduction-valence band mixing, characterized by the parameter $B$ of Kane Hamiltonian \eqref{eqHKane4}, is reflected in  \eqref{eq:eigQFKP4x4} by the pair of eigenvalues $\lambda'_1$, $\lambda'_5$, which are both determined by the whole set of the principal Hamiltonian parameters $N, M, L, A', B$. 
Note that, if one removes the mixing by setting $B=0$, this property disappears and the eigenvalues $\lambda'_4$, $\lambda'_5$  are turned into
\[
\lambda'_{04} = \E + M'+\frac{1}{2}N', \quad \lambda'_{05}  = \E +  A'. 
\]  
\subsection{Ellipticity analysis in the absence of inversion asymmetry}
We analyze the set $\lambda'_1, \lambda'_2, \lambda'_3, \lambda'_{04}, \lambda'_{05}$ associated with $B=0$ in $H^K$ 
first.
The fifth eigenvalue $\lambda'_{05}$ in \eqref{eq:eigQFKP4x4} is related to the conduction band of \eqref{eqHKane4} because its corresponding three-dimensional eigenspace ($\lambda'_{05}$ is triple degenerate) has only 3 first coordinates not equal to zero. Hence this eigenspace is orthogonal to the space associated with the valence bands. 
Those are characterized by the eigenvalues $\lambda'_1$, $\lambda'_2$, $\lambda'_3$, $\lambda'_{04}$ with degeneracy 1, 2, 3, 3, respectively.   
The following system of inequalities ensures ellipticity of $8\times8$ ZB Hamiltonian\cite{kp8ProperKane_Bahder90} with zero $B$  
\begin{equation}\label{eiginKaneN}
    \left\{
    \begin{array}{rc}
      \E + L' + N' & <0 \\[2pt]
      \E + L'-\frac{1}{2}N' & <0 \\[2pt]
      \E + M' -\frac{1}{2}N' & <0 \\[2pt]
      \E + M' + \frac{1}{2}N' & <0 \\[2pt]
      \E + A' & >0. 
    \end{array}\right.
\end{equation}
As we mentioned, the eigenvalues $\lambda'_1, \lambda'_2, \lambda'_3, \lambda'_{04}$ are related to the valence band, hence the sign of the first four innequalities from  \eqref{eiginKaneN} is the same as in \eqref{eqineig}. 
The opposite sign of the fifth inequality reflects its correspondence to the conduction band. 
Due to the electron-hole duality, the conduction band eigen-energies need to be semi-bounded from below.
The presence summand $E$ in system \eqref{eiginKaneN} is connected with the differences in the definition of Dresselhaus parameters\cite{Dresselhaus1955} and $L', M'$\cite{kp8ProperKane_Bahder90}.

To compare the result for the $8\times8$ ZB Hamiltonian with the previously obtained results for the $6\times6$ Hamiltonian we define the dimensionless parameters $\gamma'_1, \gamma'_2, \gamma_3'$ similar to the Luttinger triplet \cite{Pidgeon1966,VoonWillatzen2009,Birner2014} 
\begin{eqnarray*}
  \gamma'_{1} &=& - \frac{1}{3}(L'+2M')\frac{2m_{0}}{\hbar^2} - 1 \\
  \gamma'_{2} &=& -\frac{1}{6}(L'-M')\frac{2m_{0}}{\hbar^2} \\
  \gamma'_{3} &=&  -\frac{1}{6}N'\frac{2m_{0}}{\hbar^2}.
\end{eqnarray*}
Hereby, the system (\ref{eiginKaneN}) is transformed to
\begin{equation}\label{condKane_gamma_B_zero}
    \left\{
    \begin{array}{rc}
        -\gamma'_1 - 4\gamma'_2 - 6\gamma'_3 &< 0\\[2pt]
        -\gamma'_1 - 4\gamma'_2 + 3\gamma'_3 &< 0\\[2pt]
        -\gamma'_1 + 2\gamma'_2 + 3\gamma'_3 &< 0\\[2pt]
        -\gamma'_1 + 2\gamma'_2 - 3\gamma'_3 &< 0\\[2pt]
        1+A &> 0, 
    \end{array}\right.
\end{equation}
with $A = A'\slash E$.

The modified and the original Luttinger parameters $\gamma_1, \gamma_2, \gamma_3$ are connected by the formulas \cite{Pidgeon1966}
\begin{equation}\label{eqConvLLm}
\gamma'_1 = \gamma_1 -\frac{E_p}{3 E_g},\ 
\gamma'_2 = \gamma_2 -\frac{E_p}{6 E_g},\ 
\gamma'_3 = \gamma_3 -\frac{E_P}{6 E_g},
\end{equation}
where $E_p = P_0^2\slash E$, $E_q = E_c-E_v$ is a fundamental bandgap energy,  $P_0$ is the Kane parameter from \eqref{eqHKane4}.  

As it was expected, four out of five obtained inequalities \eqref{condKane_gamma_B_zero}, which represent the ellipticity constrains for the valence band part of $H^K$, have the structure equivalent to that for the LK Hamiltonian (\ref{eig_LK}). 
Hence, the feasibility region of the valence-band part of $H^K$ in the space of parameters  $(\gamma'_1, \gamma'_2, \gamma'_3)$ coincides with the feasibility region  $\Lambda_{-}$ of $H^{LK}$, depicted in FIG. \ref{fig:DLK}.
It means that if  $(\gamma'_1, \gamma'_2, \gamma'_3) \in \Lambda_{-}$, the valence-band part of the Hamiltonian $H^K$ in the position representation is an elliptic partial differential operator.
Then, the transformation given by \eqref{eqConvLLm} can be geometrically interpreted as a shift in the space of parameters proportional to vector $\vec{v'} = (-2,-1,-1)$.  
This shift reduces the value of $\lambda'_3$ and, as we shall soon see, brings the majority of the non-elliptic parameter triplets $(\gamma_1, \gamma_2, \gamma_3)$ closer to the feasibility region. 

The dimensionless parameter $A$ from the fifth inequality, that complements a set of ellipticity constrains \eqref{condKane_gamma_B_zero}, is responsible for a coupling between the conduction band and other states. 
It is commonly assumed that the in-basis valence bands are the major contributors to $A$.
The value of  $A$ is determined by matching its value to the effective mass of conduction band $m_c$, determined experimentally using the formula 
\begin{equation}\label{eq:A'_exp}
A = \frac{m_0}{m_c} - 1 - E_p\frac{E_g+\frac{2}{3}\Delta}{E_g(E_g+\Delta)}.
\end{equation}
The magnitude of this parameter is clearly affected by the size of band-gap $E_g$ and spin-splitting $\Delta$.
The experimental nature of $m_c$ does not factor out other possible contributions to $A$. 
For that reason we extended the collection of parameter sets from Table \ref{tabMPZBL6x6} by those stemming from the same sets of Luttinger parameters and the different values of bandgap energy $E_g$ (measured within different experimental setups). 
We also added a parameter set obtained by fitting the bandstructure of $8 \times 8$  Hamiltonian to the bandstructure calculated by ab-initio methods\cite{Bastos2016}. 
All the data pertaining to the ellipticity analysis of $8 \times 8$ Hamiltonian is collected in Table \ref{tabMPZBL8x8}. 
In each case the modified Luttinger parameters $\gamma'_1, \gamma'_2, \gamma_3'$ were calculated by using \eqref{eqConvLLm} and the values of $P_0^2$, $E_g$ provided in the dataset source. 
For those sources from the table that have $P_0^2$ unavailable we use the values collected by I.~Vurgaftman, J.~R.~ Meyer and L.~R~Ram-Mohan\cite{Vurgaftman2001}.
\sisetup{
	round-mode      = places,
	round-precision = 2,
	table-figures-integer  = 2,
	table-figures-decimal  = 2,
	table-align-text-post = false,
}

\newcounter{zbktabno}
\begin{table*}
	\centering \caption{The material data for $8\times8$ ZB Hamiltonian with $B=0$, $d$ -- distance from
		the point $(\gamma_{1},\gamma_{2},\gamma_{3})$ to the feasibility region $\Lambda_{-}$. 
		The positive values of $\lambda'_1/E -\lambda'_{05}/E$ are printed in bold.}
	\label{tabMPZBL8x8} 
	\tabcolsep=0.4em 
\begin{threeparttable}
\csvloop
{
	file=ZB_8x8_gamma_lambda_d_rho.csv,
	no head, column count=21, column names reset,
	column names={2=\one, 3=\two, 4=\three, 5=\four,6=\five, 7=\six, 8=\seven, 9=\eight,10=\nine, 11=\ten,12=\eleven, 13=\twelve, 14=\thirteen, 15=\fourteen, 16=\fifteen, 17=\sixteen, 18=\seventeen, 19=\eighteen, 20=\nineteen, 21=\twenty},
	before reading=\centering\sisetup{table-number-alignment=center, detect-weight=true, detect-inline-weight = text},
	tabular={r|l|SS[table-format=1.2]S[round-precision = 3, table-format=0.3]SS[table-format=2.2]SS|SSSSSS[table-format=1.2]S[table-format=1.2]|S[table-format=1.2]S[table-format=0.2]},
	table head=\toprule[0.3pt]\midrule[0.3pt]\bottomrule %
		\# & El %
		& {$E_p$} & {$E_g$} & {$\Delta_{SO}$} & {$A'$} & {$\gamma'_1$} & {$\gamma'_1$} & {$\gamma'_3$} %
		& {$\lambda'_1/E$} & {$\lambda'_2/E$} & {$\lambda'_3/E$} & {$\lambda'_{04}/E$} & {$\lambda'_{05}/E$} %
		& {$d$} & {$\rho$} & {$\Delta_{05}^{\text{min}}$} & {$\Delta_{05}^{\text{max}}$}
		\\\hline\midrule\bottomrule, 
	before line={\refstepcounter{zbktabno}},
	command={\thezbktabno\nineteen} & \one\text{\eighteen} %
	& \two & \three & \four & \five & \six & \seven & \eight%
	& \ifdimgreater{0 mm}{\nine mm}{\nine}{\bfseries \nine} %
	& \ifdimgreater{0 mm}{\ten mm}{\ten}{\bfseries \ten} %
	& \ifdimgreater{0 mm}{\eleven mm}{\eleven}{\bfseries \eleven} %
	& \ifdimgreater{0 mm}{\twelve mm}{\twelve}{\bfseries \twelve} %
	& \ifdimgreater{0 mm}{\thirteen mm}{\thirteen}{\bfseries \thirteen} %
	& \fourteen & \fifteen & \sixteen & \seventeen,
	table foot=\toprule\midrule[0.3pt]\bottomrule[0.3pt],
	filter = \(\thecsvinputline>1\) \and \(\thecsvinputline<53\)}
\begin{tablenotes}\footnotesize
	\item [a] Set 1 from \onlinecite{LB1}
	\item [b] Set 2 from \onlinecite{LB1}
	\item [c] Set 3 from \onlinecite{LB1}
	\item [d] Set 4 from \onlinecite{LB1}
	\item [e] Set 5 from \onlinecite{LB1}
	\item [f] Set 6 from \onlinecite{LB1}
	\item [g] Obtained by extrapolation from 5-level model
	\item [h] Measured at $T = 300K$
\end{tablenotes}
\end{threeparttable}
\end{table*}

The ellipticity conditions of $H^K$ are still violated for all materials presented in Table
\ref{tabMPZBL6x6}. The situation is, however,  more complex than for the $6 \times 6$ Hamiltonian. 
To illustrate that, we supplied in Table \ref{tabMPZBL8x8} the values of $\lambda'_1 -\lambda'_{05}$, the distance $d$ to the feasibility region from $(\gamma'_1, \gamma'_2, \gamma'_3)$ and the measure of non-ellipticity $\rho$ which is defined in the same way as for the $6 \times 6$ Hamiltonian case. 

Overall, we can confirm the reduction of average distance to the feasibility region for all materials, especially for InAs and InSb. 
Furthermore, for several materials there exist parameter sets that are close to satisfy the full set of ellipticity constraints described by \eqref{condKane_gamma_B_zero}. 
Those are narrow gap semiconductor InSb (sets \#\ref{mp:ZBK:InSb_LB1_1}; \#\ref{mp:ZBK:InSb_LB1_2} from Table \ref{tabMPZBL8x8})  and,  perhaps more surprisingly, the materials with larger band-gap InP, AlAs and AlSb  ( sets \#\ref{mp:ZBK:InP_LB1_2}, \#\ref{mp:ZBK:AlAs_LB1};\#\ref{mp:ZBK:AlAs_Mad}, and \#\ref{mp:ZBK:AlSb_Vurg1}). 
For these materials the corresponding parameter sets can be made elliptic by direct adjustment of $\gamma'_1, \gamma'_2, \gamma'_3, A$.

Certain parameter sets for AlP, AlSb and InAs satisfy ellipticity conditions for the valence band part of the Hamitonian and do not satisfy the conduction-band constraint (inequality 5 from \eqref{condKane_gamma_B_zero}). 
Among those, the sets \#\ref{mp:ZBK:AlP_Vurg1}, \#\ref{mp:ZBK:AlSb_Vurg1} for AlP, AlSb  reported in [\onlinecite{Vurgaftman2001}] differ sharply in the size of $\gamma_2$ from two other sets for these materials collected in Table \ref{tabMPZBL8x8}.
For AlP this can explained by the fact that in the absence of direct experimental data most of the material parameters were extrapolated from measurements for ternary alloys and ab-initio calculations which carries a lot of uncertainty.  
The authors of [\onlinecite{Vurgaftman2001}] performed readjustment of the Luttinger parameters to better match the experimental photoluminescence results on AlP$/$GaP heterostructures\cite{Issiki1995}. 
The set \#\ref{mp:ZBK:AlSb_Vurg1} for AlSb is based on the available theoretical  calculations from various sources and the simultaneous fitting of $\gamma_1, \gamma_2, \gamma_3$ to the experimentally-determined hole effective masses along [001], [110] and [111] directions \cite{Vurgaftman2001}. 

For InAs we can judge from the size of $\lambda'_1 -\lambda'_{04}$ that the triplets $(\gamma'_1, \gamma'_2, \gamma'_3)$ of its parameter sets are right near the side of $\Lambda_{-}$ described by $\lambda'_{2}=0$. Two 
are inside (sets \#\ref{mp:ZBK:InAs_Vurg1};\#\ref{mp:ZBK:InAs_LB1_1}) and two others 
are slightly off (sets \#\ref{mp:ZBK:InAs_LB1_2};\#\ref{mp:ZBK:InAs_Mad}). 
The values of $\lambda'_{05}$ for all four parameter sets are grouped near the value $\lambda'_{05} = -4.8$ and thus the conduction part of the Hamiltonian is, again, far from being elliptic. 

All material parameter sets for GaAs violate two out of four ellipticity conditions for the valence-band part,  although one parameter set  \#\ref{mp:ZBK:GaAs_LB1_2} from Table \ref{tabMPZBL8x8} stays close to  $\Lambda_{-}$ ($d=\num{0.341226582094873}$).  
However, it violates the ellipticity condition for the conduction-band part by the same margin of approximately $-2.9$ as do other sets of GaAs parameters, let alone set \#\ref{mp:ZBK:GaAs_Bastos} where the margin is slightly lower: $\lambda'_{05} \approx -2.34$. 
The indicated reduction of margin should be attributed to the optimization procedure\cite{Bastos2016} used to acquire set \#\ref{mp:ZBK:GaAs_Bastos}.
As far as the ellipticity is concerned, this optimization procedure is no more effective than other acquisition methods.

The sets for GaN and AlN are failing first two valence-part constraints from \eqref{condKane_gamma_B_zero} just like the most of the other material parameters. 
One exception is the set number \ref{mp:ZBK:GaN_Cardona} for GaN\cite{Cardona_05}, where the spherical symmetry of the heavy-hole and the light-hole bands is assumed. 
This assumption leads to the larger values of $\gamma_1, \gamma_3$ and smaller $\gamma_2$; and, as a consequence, more than five times smaller ratio $\rho$ between positive and negative eigenvalues.

Even a more severe situation is observed for InN. The conduction-band eigenvalue $\lambda'_{05}$ is noticeably below zero for all three available datasets ($\lambda'_{05} \approx \num{-4.54002041371321}$; $\num{-7.71874975696631}$; $\num{-2.87312316725118}$ for sets \#\ref{mp:ZBK:InN_Vurg1}; \#\ref{mp:ZBK:InN_Vurg2}; \#\ref{mp:ZBK:InN_Rinke} accordingly).  
In addition, three out of four valence-band conditions are violated. 
It is important to highlight that the recently obtained set of parameters \#\ref{mp:ZBK:InN_Rinke} features roughly two times larger values of $\gamma_1, \gamma_2$ and $\gamma_3$ and noticeably smaller $E_p, E_g$ than two other sets \#\ref{mp:ZBK:InN_Vurg1}, \#\ref{mp:ZBK:InN_Vurg2} reported earlier\cite{Vurgaftman2001, Vurgaftman2003}. 
As demonstrated in [\onlinecite{Rinke2008}] set \#\ref{mp:ZBK:InN_Rinke} recovers bandstructure better than two others sets for InN, discussed above.
In terms of ellipticity, this set results in a lower, than others, distance to $\Lambda_{-}$  ($d \approx\num{0.842567107552504}$) and lower  $\lambda'_{05} \approx \num{0.169245283018868}$.  
 
\begin{figure}
	\centering
	\includegraphics[width=1\linewidth]{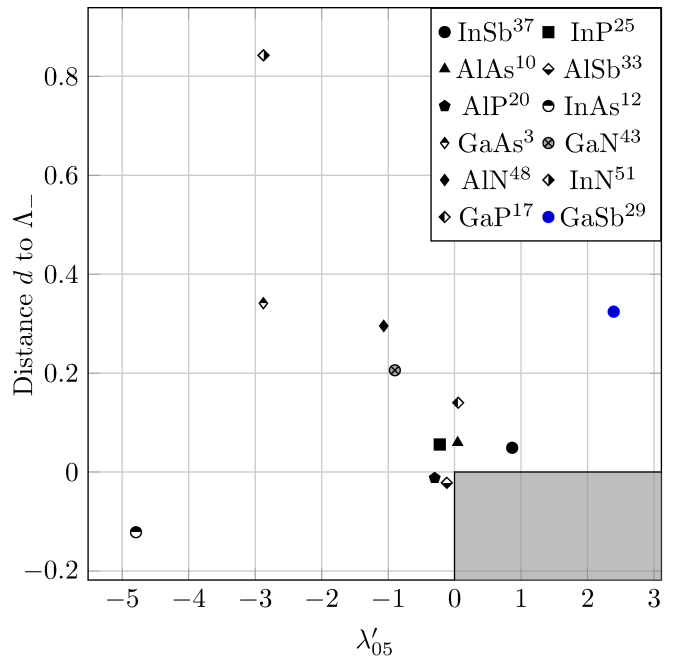}
	\caption{The values of quantities $\lambda'_{05}, d$ for the list of selected material parameter sets from Table \ref{tabMPZBL8x8} (color online).  The shaded region indicates pairs $(\lambda'_{05}, d)$ based on the elliptic parameters of $8 \times 8$ model\cite{kp8ProperKane_Bahder90} }
	\label{fig:ZBK_sel_mat_par_d_lambda05}
\end{figure}

For GaP and GaSb the data seem inconclusive as the size and the sign of eigenvalues \eqref{eq:eigQFKP4x4} are dependant on the choice the material parameter dataset. 
Sets \#\ref{mp:ZBK:GaP_LB1_1}, \#\ref{mp:ZBK:GaSb_LB1_1} from Landolt-B\"ornstein\cite{LB1}, based on the earlier data of P. Lawaetz\cite{kp7Lawaetz71}, are most favourable in terms of ellipticity:  $d \approx\num{0.140059606511186}$, $\lambda'_{05} \approx \num{0.053418803418803}$ for GaP;   $d \approx\num{0.324200259259466}$, $\lambda'_{05} \approx \num{2.39227265902608}$ for GaSb.  
As a summary of the above analysis, we visualize in FIG. \ref{fig:ZBK_sel_mat_par_d_lambda05} the values of $ \lambda'_{05}, d$ for the selected parameter sets with the material-wise minimal distance to $\Lambda_{-}$. 
In this figure the ellipticity of Hamiltonian is depicted by the region (shaded in gray) where $\lambda'_{05} > 0$ and $d < 0$ simultaneously. 

It is worth noticing that roughly $76 \si{\percent}$ of parameter sets for analyzed materials fail the conduction-band constraint $\lambda'_{05}>0$. 
This group includes all datasets for  GaAs, InAs, AlP, AlSb, GaN, AlN, InN, quite important for applications.
The positive (negative) sign of $\lambda'_{05}$ is responsible for positive(negative) gain in the energy as we go from one conduction-band eigenvalue of Hamiltonian to the next in the position representation. 
In the momentum representation, the eigenvalue's sign and its magnitude is responsible for upward (downward) curvature of conduction band. 
In addition to the highlighted in section \ref{sec6x6} issues caused by the non-ellipticity of valence-band part of the Hamiltonian $H^K$, the violation of condition $\lambda'_{05}>0$ entails the existence of conduction-band related eigenstates of $H^K$ with energies in the band-gap or the regions related to valence bands\cite{Veprek07,Veprek08,foreman_sp_07,Birner2014}. 
This obviously poses a serious problem in applications.

A rescaling procedure was introduced by B.~Foreman in [\onlinecite{Foreman_sp_97}] (see also the work of S.~Birner [\onlinecite{Birner2014}]) and has been adopted\cite{Vinas2017, Birner2007} ever since as a way to make $\lambda'_{05}$ positive and avoid the above-described type of spurious solutions.
The idea of the procedure is to adjust the momentum matrix element $E_p$ so that $\lambda'_{05}$ is no longer negative. 
Let us assume that we target some value of $\lambda'_{05}=a$. 
Then, by using  the definition of $\lambda'_{05}$ and \eqref{eq:A'_exp}, we obtain\cite{Birner2014} 
\begin{equation}\label{eq:Kane_Ep_adj}
E_p = \left (\frac{m_0}{m_c} - a\right ) \frac{E_g(E_g+\Delta_{SO})}{E_g+\frac{2}{3}\Delta_{SO}}.
\end{equation}
Two values $a=0$ and $a=1$ are considered in the literature as a target for rescaling. 
The new value of $E_p$ will necessary affect the values of modified Luttinger parameters $\gamma'_1, \gamma'_2, \gamma'_3$ which are defined by \eqref{eqConvLLm}. 
To figure out how this procedure would impact the ellipticity of the entire $8\times8$ ZB Hamiltonian $H^K$ one needs to rewrite eigenvalues $\lambda'_1 -\lambda'_{04}$ as functions of $E_p,E_g$: 
\begin{equation*}
\lambda'_1 = \lambda_1 + 2E \frac{E_p}{E_g}, \  
\lambda'_{2/04} = \lambda_{2/4} + \frac{E}{2} \frac{E_p}{E_g},  \ 
\lambda'_3 = \lambda_3 - \frac{E}{2} \frac{E_p}{E_g}.
\end{equation*}
By combining the above representations with \eqref{eq:Kane_Ep_adj} we obtain a new version of ellipticity constraints \eqref{condKane_gamma_B_zero} for the valence-band part of $H^K$ 
\begin{equation}
\label{eq:Kane_eig_a}
\begin{aligned}
\lambda_1 + 2 E_m - \frac{2 a} {E_r} < 0, \quad  
\lambda_2 + \frac{1}{2} E_m - \frac{a}{2E_r} < 0, \\
\lambda_3 - \frac{1}{2} E_m + \frac{a}{2E_r}< 0, \quad
\lambda_4 + \frac{1}{2} E_m - \frac{a}{2E_r}< 0, 
\end{aligned}
\end{equation}
where $E_m = \dfrac{m_0}{m_c E_r}$, $E_{r} = \dfrac{E_g+\frac{2}{3}\Delta_{SO}}{E(E_g+\Delta_{SO})}$ are two material-dependent constants. 
Now we substitute back $a = \lambda'_{05} + \Delta_{05}$ and solve system of inequalities \eqref{eq:Kane_eig_a} with respect to $\Delta_{05}$. 
As a result, it will give us the range for the values of the rescaling parameter $\Delta_{05}$ that make the valence-band part of Kane Hamiltonian elliptic
\begin{equation}\label{eq:Kane_range_delta}
E_r\max\left \{ \frac{1}{2}\lambda'_1, 2\lambda'_2, 2\lambda'_{04}\right \}
< \Delta_{05} <
-2 \lambda'_3 E_r. 
\end{equation}
The calculated values for the ranges from \eqref{eq:Kane_range_delta}  are provided in the last two columns of Table \ref{tabMPZBL8x8}. 
If $-\lambda'_{05}<-2 \lambda'_3 E_r$, then the Hamiltonian can be made elliptic by setting $\Delta_{05}$ to the arbitrary value within range \eqref{eq:Kane_range_delta} so that $\lambda'_{05} + \Delta_{05}$ is positive.
In practice one would also like to make sure that the  numerical  inaccuracies introduced by the eigenvalue calculation procedure for $H^K$ will not overturn any of the signs of $\lambda'_1$ -- $\lambda'_{05}$.
To minimize that possibility and to keep $\Delta_{05}$ reasonably small we suggest the following formula for the selection of $\Delta_{05}$
\begin{equation}\label{eq:ZBK_delta05_sel}
\Delta_{05} =
\begin{cases}
2\Delta_{m}+0.1, \quad  &\Delta_{m}+\lambda'_3 E_r  < -0.1,\\
\Delta_{m}- \lambda'_3 E_r, \quad &\text{otherwise}.
\end{cases}
\end{equation}
with
$\Delta_{m} = \max\left \{ \frac{1}{4}\lambda'_1 E_r, \lambda'_2 E_r, \lambda'_{04} E_r,- \dfrac{\lambda'_{05}}{2E} \right \}$.

We carried out the rescaling procedure for the material parameters from Table \ref{tabMPZBL8x8} and selected the sets with minimal $\Delta_{05}$ for every given material. 
The resulting values of readjusted $E_p$, $A'$ along with new values of modified Luttinger parameters are presented in Table  \ref{tabMPZBL8x8resc}. 
It is also  worth noting that the resulting value of $1+A$ in our case is never equal to zero or one, as it was usually assumed by authors before\cite{Foreman_sp_97, Birner2014}. 
For many materials the value of adjusted parameter $A$ is greater than 0. 

\begin{figure}[ht]
\begin{overpic}[width=0.2\textwidth]{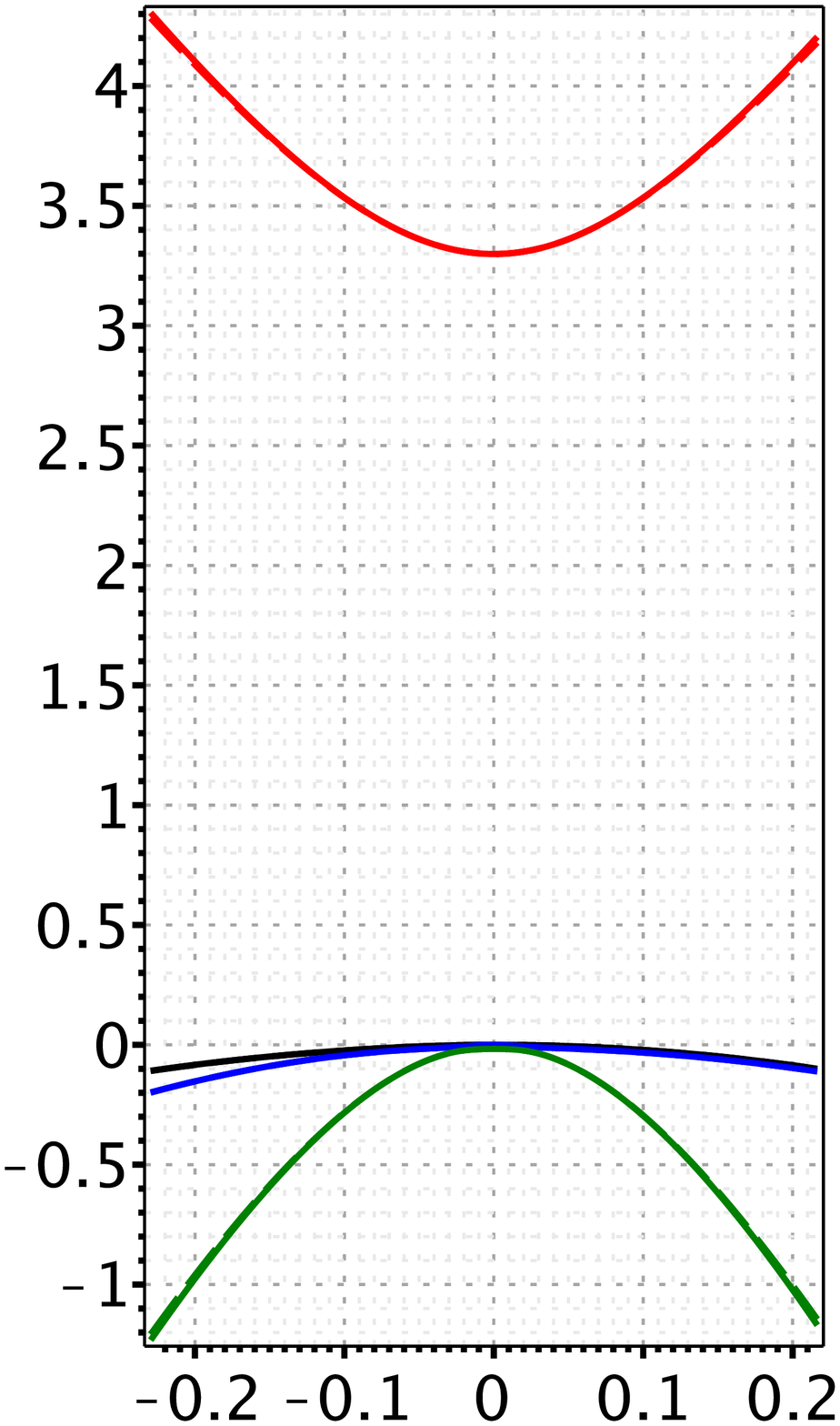}
	\put (1,50) {\rotatebox{90}{\small Energy (eV)}}
	\put (12,2) {\small Fraction of the path}
	\put (20.5,12) {$ K \leftarrow \Gamma \rightarrow L$}
	\put (29,93) {\large GaN}
	\put (3,93) {\large a)}
	\thicklines
	\put (15.5, 55){%
	\colorbox{White}{\fbox{\parbox{6em}{\tiny%
	\begin{tabular}{ll}
		Rescaled & Original\\ 
		{\color{Red}\lrsolid} \hfill CB &
		{\color{Red}\lrdotdashed} \hfill CB
		\\%
		{\color{Black}\lrsolid} \hfill HH &%
		{\color{Black}\lrdotdashed} \hfill HH
		\\%
		{\color{Blue}\lrsolid} \hfill LH &%
		{\color{Blue}\lrdotdashed} \hfill LH
		\\%
		{\color{OliveGreen}\lrsolid} \hfill SO &%
		{\color{OliveGreen}\lrdotdashed} \hfill SO
	\end{tabular}		
	}}}
}

\end{overpic}
\raisebox{4.5em}{
\begin{overpic}[width=0.25\textwidth]{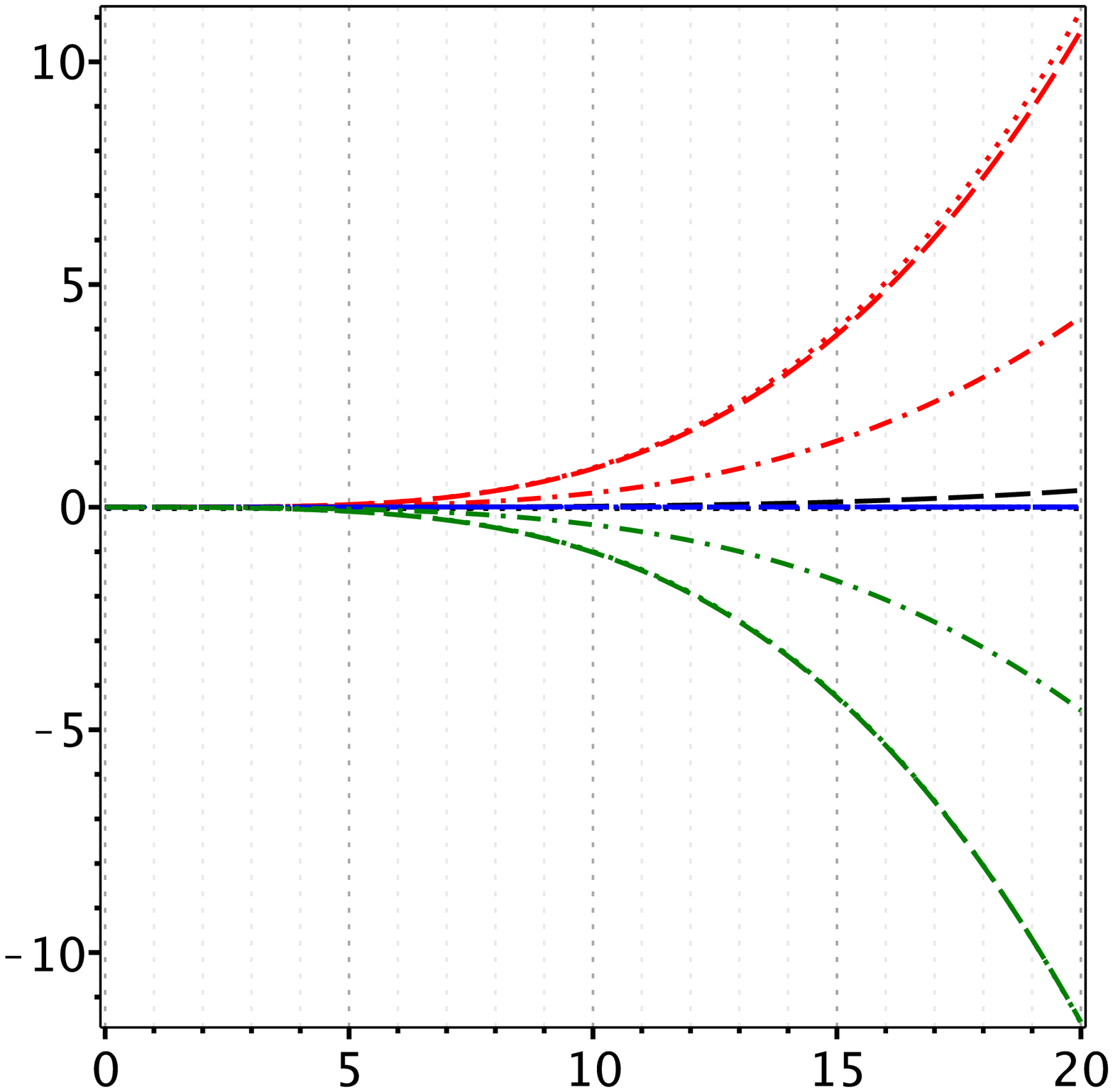}
	\put (1,30) {\rotatebox{90}{\small Energy (meV)}}
	\put (25,2) {\small Percent of the path}
	\put (45,90) {\large GaN}
	\put (14,12) {$ \Gamma$}
	\put (-3,91) {\large b)}
	\thicklines
	\put (15.6,-8) {\small CB: 
		{\color{red}\lsdotted} \hspace{0.4em} $ \Gamma L$, %
		{\color{red}\lsdashed} \hspace{0.5em} ${ \Gamma K}$, %
		{\color{red}\lsdotdashed} \hspace{0.5em} ${ \Gamma X}$%
	    }
	\put (15.3,-15) {\small HH: %
	{\color{black}\lsdotted} \hspace{0.4em} $ \Gamma L$, %
	{\color{black}\lsdashed} \hspace{0.5em} ${ \Gamma K}$, %
	{\color{black}\lsdotdashed} \hspace{0.5em} ${ \Gamma X}$%
	}
	\put (16,-22) {\small LH: %
	{\color{blue}\lsdotted} \hspace{0.4em} $ \Gamma L$, %
	{\color{blue}\lsdashed} \hspace{0.5em} ${ \Gamma K}$, %
	{\color{blue}\lsdotdashed} \hspace{0.5em} ${ \Gamma X}$%
	}
	\put (16,-29) {\small LH: %
	{\color{OliveGreen}\lsdotted} \hspace{0.4em} $ \Gamma L$, %
	{\color{OliveGreen}\lsdashed} \hspace{0.5em} ${ \Gamma K}$, %
	{\color{OliveGreen}\lsdotdashed} \hspace{0.5em} ${ \Gamma X}$%
	}
\end{overpic}
}
\caption{Comparison of original and rescaled parameter sets for GaN (color online): Conduction band (CB), heavy-hole (HL), light-hole (LH) and the split-off band (SO). a)  Bandstructure along the fraction of symmetry path $K-\Gamma-L$: original set \ref{mp:ZBK:GaN_LB1} (solid), rescaled set from Table \ref{tabMPZBL8x8resc} (dashed); b) bandstructure adjustment error along the paths $ \Gamma L$, ${ \Gamma K}$ and ${ \Gamma X}$ }
\label{fig:ZBK_resc_bs_comp:GaN}
\end{figure}

\begin{figure*}[ht]
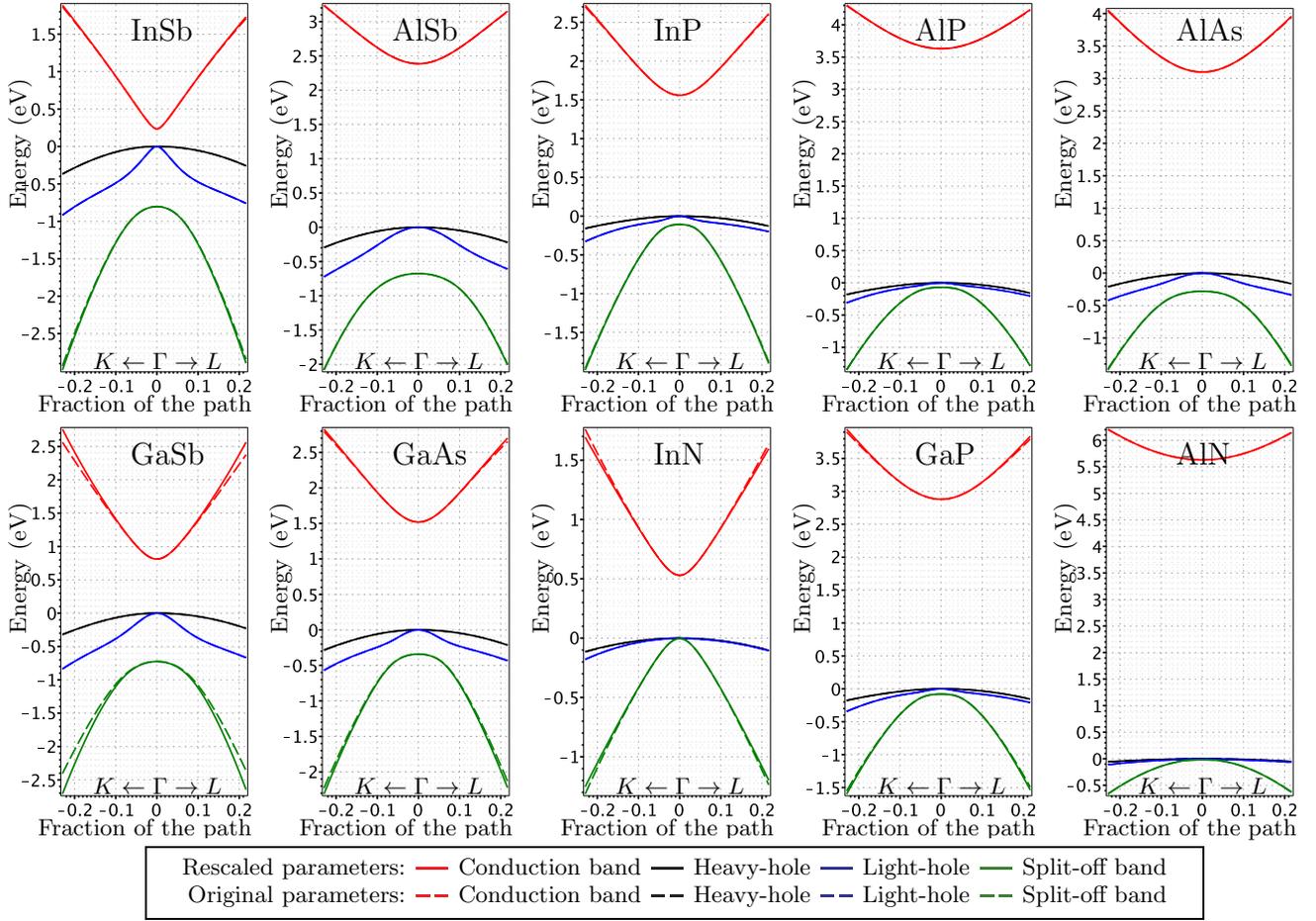

	\begin{center}
		\catcode`"=9	
		\csvreader[column count=7, head, column names reset, %
		column names={1=\one, 2=\two, 3=\three, 4=\four,5=\five, 6=\six, 7=\seven, 8=\eight}, %
		filter expr ={ test{\ifcsvnotstrcmp{\two}{GaN}}}]
		{bs_errors_resc_sort_rel_diff_ep.csv}{}{%
			\begin{overpic}[width=0.19\textwidth]{\seven}
				\put (1,50) {\rotatebox{90}{\small Energy (eV)}}
				\put (8,2) {\small Fraction of the path}
				\put (21,12) {$ K \leftarrow \Gamma \rightarrow L$}
				\put (30,90) {\large \two}
			\end{overpic} %
		}\\
	\end{center}
	\vspace*{-1.4em}\hspace{1.4em}
	\colorbox{White}{\fbox{\parbox{43em}{\small%
				\begin{tabular}{rllllllll}
					Rescaled parameters: &%
					{\color{Red}\lrsolid} & Conduction band &%
					{\color{Black}\lrsolid} & Heavy-hole &%
					{\color{Blue}\lrsolid} & Light-hole &%
					{\color{OliveGreen}\lrsolid} & Split-off band   
					\\%
					Original parameters: &%
					{\color{Red}\lrdashed} & Conduction band &%
					{\color{Black}\lrdashed} & Heavy-hole &%
					{\color{Blue}\lrdashed} & Light-hole &%
					{\color{OliveGreen}\lrdashed} & Split-off band
				\end{tabular}		
	}}}
	\caption{Comparison of bandstucture for the selected original parameter sets from Table \ref{tabMPZBL8x8} and the rescaled sets from Table \ref{tabMPZBL8x8resc} (color online).%
		The band dispersion is plotted from even eigenenergies of $8\times 8$ Hamiltonian\cite{kp8ProperKane_Bahder90} along the fraction of symmetry path $K-\Gamma-L$ in the vicinity of $ \Gamma$: original sets (solid line), rescaled sets (dashed line).}
	\label{fig:ZBK_resc_bs_comp}
\end{figure*}

To see the impact of rescaling on the band dispersion,  in Table  \ref{tabMPZBL8x8resc} we also supplied a maximum absolute difference~(adjustment error) between the corresponding bands of bandstructure calculated over the 20\% of three high symmetry paths  $ \Gamma L$, ${ \Gamma K}$ and ${ \Gamma X}$ pertaining to the first Brillouin zone (FBZ).
Such a size of the domain for comparison is common\cite{Vurgaftman2001} and motivated by the existing evidence\cite{Bastos2016} that the accurate fit of the $k \cdot p $ bandstructure to the state-of-the-art ab-initio calculations is possible over this part of FBZ.
To ascertain the band that contributes most to the error we supplied in FIG. \ref {fig:ZBK_resc_bs_comp:GaN} a), and FIG. \ref {fig:ZBK_resc_bs_comp} the graphical comparison of band-structure diagrams for every set from Table \ref{tabMPZBL8x8resc} and the original sets of material parameters from Table \ref{tabMPZBL8x8}, on which they are based. 
For clarity, only bands with even numbers in the representation of $8\times 8$ Hamiltonian\cite{kp8ProperKane_Bahder90} are plotted in these figures.  

\sisetup{
	round-mode      = places,
	round-precision = 3,
	table-number-alignment = left,
	table-figures-integer  = 1,
	table-figures-decimal  = 3,
	table-align-text-post = false,
}

\begin{table}
	\centering \caption{Selected material parameters, rescaled via \eqref{eq:ZBK_delta05_sel} together with the difference between corresponding bands of $8 \times 8$ Hamiltonian\cite{kp8ProperKane_Bahder90} for original and rescaled parameters}
	\label{tabMPZBL8x8resc} 
	\tabcolsep=0.5em 
	\begin{threeparttable}
		\csvloop
		{
			file=ZB_8x8_B_0_selected_rescaled_sort_bs_error.csv,
			no head, column count=17, column names reset,
			column names={1=\one, 2=\two, 3=\three, 4=\four,5=\five, 6=\six, 7=\seven, 8=\eight,9=\nine, 10=\ten,11=\eleven,12=\twelve, 13=\thirteen
			},
			tabular={@{\hskip 2pt}lS[round-precision = 2, table-format=1.2]S[round-precision = 2, table-format=1.2]S[table-format=1.2]SS[table-format=1.2]S[table-format=1.2]S[round-precision = 2, table-format=2.1]S[round-precision = 2, table-format=3.2]@{\hskip 2pt}},
			table head=\toprule[0.3pt]\midrule[0.3pt]\bottomrule %
			El\tnote{a} & {$\Delta_{05}$\tnote{b}} %
			
			& {$E_p$} & {$A'$} & {$\gamma'_1$} & {$\gamma'_2$} & {$\gamma'_3$} %
			& {$\lambda_v$\tnote{c}} & {$err$\tnote{d}}%
			\\\hline\midrule\bottomrule, 
			command= \two\tnote{\ref{\thirteen}} %
			& \three & \four & \five & \six & \seven & \eight & \nine & \eleven, 
			table foot=\toprule\midrule[0.3pt]\bottomrule[0.3pt],
			filter = \(\thecsvinputline>1\) \and \(\thecsvinputline<53\)%
		}
		\begin{tablenotes}\footnotesize
			\item [a] Refer to the original dataset number from Table \ref{tabMPZBL8x8}
			\item [b] The quantity $\Delta_{05}$ describes the size of adjustment to $A'$
			\item [c] The values of $\lambda_v$ are calculated via $\lambda_v = \max\{\lambda'_1, \lambda'_2, \lambda'_3, \lambda'_{04}\}$ \hfill
			\item [d] Maximum difference in meV between the bandstrucure for the original parameters from Table \ref{tabMPZBL8x8} and the rescaled parameters calculated by  using \eqref{eq:ZBK_delta05_sel} over the 20 \% of the paths $ \Gamma L$, ${ \Gamma K}$ and ${ \Gamma X}$
		\end{tablenotes}
	\end{threeparttable}
\end{table}

As one immediately notices from the last column of Table \ref{tabMPZBL8x8resc}, the chosen sets for AlN, AlP, AlSb, and AlAs   materials are least susceptible to the performed rescaling procedure.
The differences between the bandstructure for modified parameters sets (four topmost rows from Table \ref{tabMPZBL8x8resc}) and the original sets for materials from this group are less than 10 meV. 
We will call these differences  banstructure adjustment errors or simply errors, when it is unambiguous.
For GaN and InP the errors of approximately 11 meV are also visually indistinguishable in FIG. \ref{fig:ZBK_resc_bs_comp:GaN} a) and FIG. \ref{fig:ZBK_resc_bs_comp}. 
Thus we supplied in FIG. \ref{fig:ZBK_resc_bs_comp:GaN} b) their plot for GaN, that has an appropriate vertical scaling. 
This plot, typical for all analyzed materials except of InN, shows the behaviour of bandstructure adjustment error along three main paths $ \Gamma L$, ${ \Gamma K}$ and ${ \Gamma X}$.    
For the GaN, the errors along directions $ \Gamma L$, ${ \Gamma K}$ are about 11 meV, while the band errors in the direction ${ \Gamma X}$ is around two times lower. 
What makes this material unique is the fact that such tiny errors are the result of a significant change in the values of material parameters during the rescaling. 
Namely, the difference in $E_p$ is around $29 \%$ and above $100 \%$  for $\gamma'_{1}$, $\gamma'_{2}$. 

The situation is more close to the anticipated for another group of materials: GaP, InN, GaAs.
The relative differences in $E_p \approx 19 \%$ for all three materials, but the error is higher for GaAs  than for GaP and InP: 50.62 meV vs 23.34 meV and 35.02 meV, respectfully.
This can be explained by a closer proximity of $p$-like conduction bands in GaAs, that are treated perturbatively in the current model.
For indium antimonide the band adjustment error of 32.66 meV (barely visible as a slightly higher curvature of conduction and SO bands in the first plot of FIG. \ref{fig:ZBK_resc_bs_comp} ) lays within the same range as for GaP, InN, GaAs.
What is unusual is that these differences in band dispersion were produced by the smallest (among all analyzed  materials) adjustment of $E_p$ -- $0.6 \%$, which resulted in only approximately $ 10 \% $ increase of $\gamma'_{1}$, $\gamma'_{3}$. 
Such error sensitivity might be attributed to the very small bandgap (see FIG. \ref{fig:ZBK_resc_bs_comp}).
The conduction band (CB) adjustment error for InSb is equal to 10.87 meV. 
That is about 3 times smaller than the valence band adjustment error and therefore invisible in the plot.
The same is true for heavy hole and light hole bands. 

Similar tendencies are valid for other materials from Table \ref{tabMPZBL8x8resc}. 
The performed adjustment of $A$ leads to a slightly noticeable change in the conduction band dispersion. 
Heavy hole~(HH) and light hole~(LH) bands remain visually unaffected even though the differences are non-zero. 
The rescaling also causes an increase in the curvature of the split-off~(SO) band, making it the main source of total valence-band adjustment error. 


The maximum adjustment error of 171.4 meV was observed in gallium antimonide. 
We postpone a detailed discussion of GaSb till the next subsection
and focus now on the following question: How the dispersion of  CB and SO band can be corrected without braking ellipticity of the $H^K$ Hamiltonian? 

\subsection{Ellipticity analysis for $8 \times 8$ ZB Hamiltonian with inversion-asymmetry present }
To answer the question posed at the end of previous subsection, we will consider here the ellipticity conditions for the case of non-zero $B$ in \eqref{eqHKane4}.
In this case the ellipticity region in the parameter space $A', B, \gamma'_1, \gamma'_2, \gamma'_3$ is described by the system of inequalities 
\begin{equation}\label{eq:inneqQFKP4x4}
\left \{
	\begin{aligned}
		\max\left \{ - 4\gamma'_2 - 6\gamma'_3, 3\gamma'_3 - 4\gamma'_2, 3\gamma'_3 + 2\gamma'_2 \right \} &< \gamma'_1\\[6pt]
		E + A' + \lambda'_{04}  -  
		\sqrt{\left (E + A' - \lambda'_{04} \right )^2 + 2 B^2} &< 0 \\[6pt]
		E + A' + \lambda'_{04}  +  \sqrt{\left (E + A' - \lambda'_{04} \right )^2 + 2 B^2} &> 0. \\[6pt]
	\end{aligned}
 \right .
\end{equation}
The first inequality is just a compact form of inequalities 1-3 from \eqref{condKane_gamma_B_zero},  the value of $\lambda'_{04}$ is equal to the one defined above, but written in a new parameter notation $\lambda'_{04} = -\gamma'_1 + 2 \gamma'_2 - 3 \gamma'_3$. 

Despite a more complicated structure than in the situation with zero $B$, discussed earlier, one out of two  $B$-dependent constrains in \eqref{eq:inneqQFKP4x4} is always fulfilled. 
To be more specific: if $E+A' \geq -\lambda'_{04}$ the third inequality from \eqref{eq:inneqQFKP4x4} is redundant, else,  the second one is redundant. 
In each case, the remaining non-redundant inequality leads to the following constraint on $B^2$
\begin{equation}\label{eq:4x4constrB}
B^2 - 2\E^2(1+A)(-\gamma'_1 + 2 \gamma'_2 - 3 \gamma'_3) > 0.
\end{equation}
The combination of \eqref{eq:4x4constrB} with the first inequality from \eqref{eq:inneqQFKP4x4} yields a system of ellipticity constraints for $8\times8$ ZB Hamiltonian\cite{kp8ProperKane_Bahder90} with non-zero $B$
\begin{equation}\label{eq:el_constrZBKane_nzB}
\left  \{
\begin{aligned}
	\max\left \{ - 4\gamma'_2 - 6\gamma'_3, 3\gamma'_3 - 4\gamma'_2, 3\gamma'_3 + 2\gamma'_2 \right \} &< \gamma'_1\\[2pt]
	 2\E^2(1+A)(-\gamma'_1 + 2 \gamma'_2 - 3 \gamma'_3) &< B^2.
\end{aligned}
\right .
\end{equation} 

Inequality \eqref{eq:4x4constrB} is fulfilled for any $B$,  when the parameter set $A, \gamma'_1, \gamma'_2, \gamma'_3$ satisfies  conditions 4-5 from \eqref{condKane_gamma_B_zero}. 
Consequently,  the admissible, in terms of \eqref{condKane_gamma_B_zero}, material parameters with zero $B$ remain admissible even after the value of $B$ is set to some nonzero number. 
In other words, $B$ can be treated as an additional fitting parameter to be used in the subsequent adjustment step after the rescaling procedure is performed, but the additional ellipticity preserving readjustment of bandstructure is needed.
We are especially interested in correcting the improper dispersion of the CB and SO bands, since it is a major source of errors (see Table \ref{tabMPZBL8x8resc}) for most of the rescaled material parameter sets. 

Conducted numerical experiments\cite{Note1} with different values of $B$ indicate that the increase in $|B|$ leads to the increase in the curvature of conduction and SO bands along $\Gamma  K$, $\Gamma  L$ and $\Gamma  X$ directions.
For InN, that makes the CB adjustment error along those directions smaller at the expense of larger difference in SO band dispersion between the original set and the rescaled parameter set with non-zero $B$.
The indicated behaviour of error and the fact that the error's dominating contribution comes from CB and SO bands (see FIG. \ref{fig:ZBK_resc_bs_comp:GaN} b) mean that there exists an optimal value $B$ that minimizes the error for chosen energy bands.
For small errors and $B>0$ the indicated behaviour is also influenced by the spin-slitting of bands away from $\Gamma$. 
Error can not be minimized for GaSb material, because the rescaled parameters with $B=0$ already yield visibly higher curvature of CB than the original parameters (see GaSb plot in FIG. \ref{fig:ZBK_resc_bs_comp}).
We performed the error minimization by adjusting $B$ for each selected parameter set from Table \ref{tabMPZBL8x8resc} and confirm that for all materials, except of InN, increase in $|B|$ makes error larger.

\begin{figure}[t]
	\hspace*{-4mm}
	\begin{overpic}[width=0.49\linewidth]{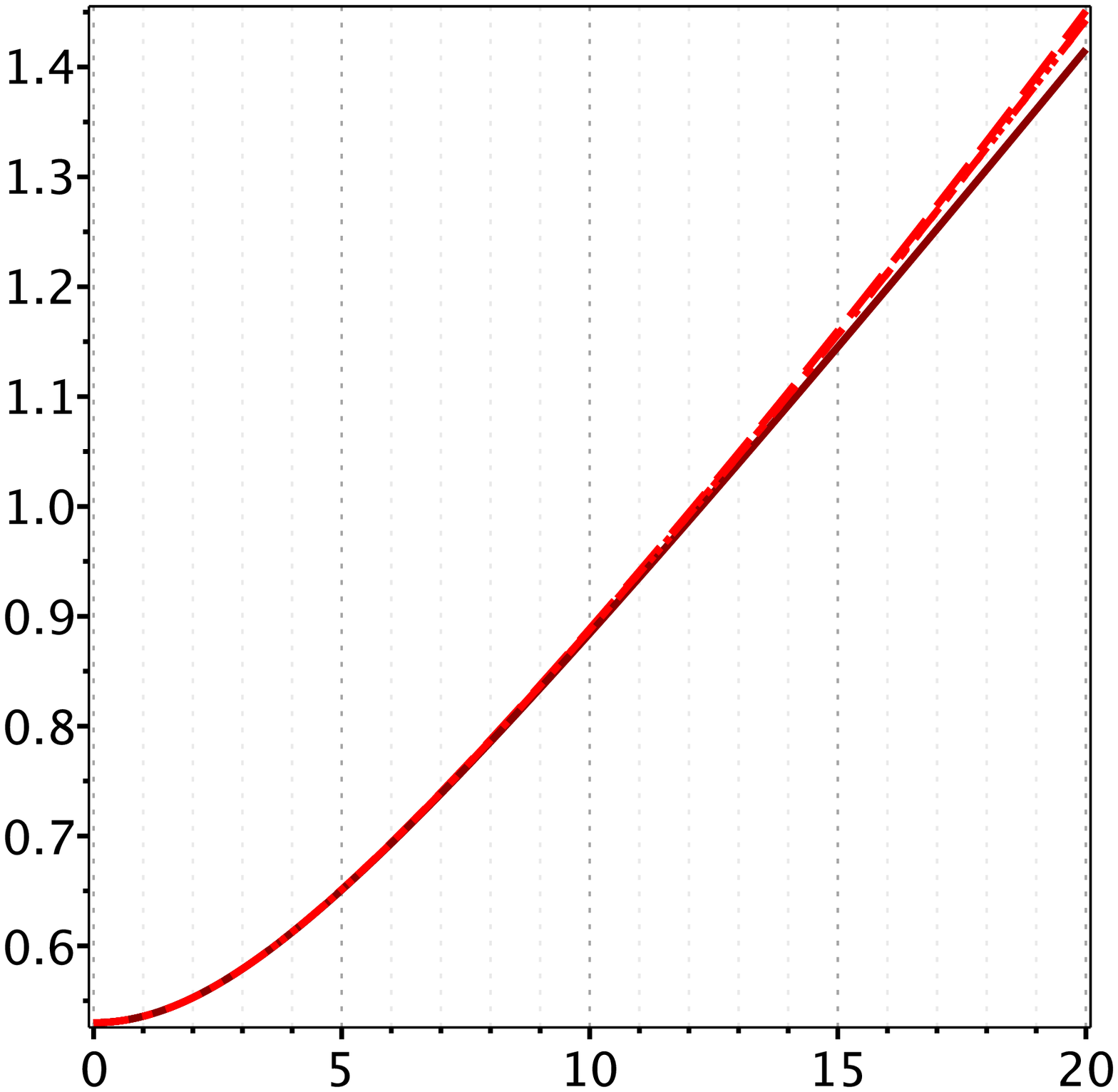}
		\put (-3,30) {\rotatebox{90}{\small Energy (eV)}}
		\put (24,2) {\small Percents of the path}
		\put (44,13) {$\Gamma \rightarrow K$}
		\put (15,85) {\large a)}
	\end{overpic}
	\hspace*{-1mm}
	\begin{overpic}[width=0.5\linewidth]{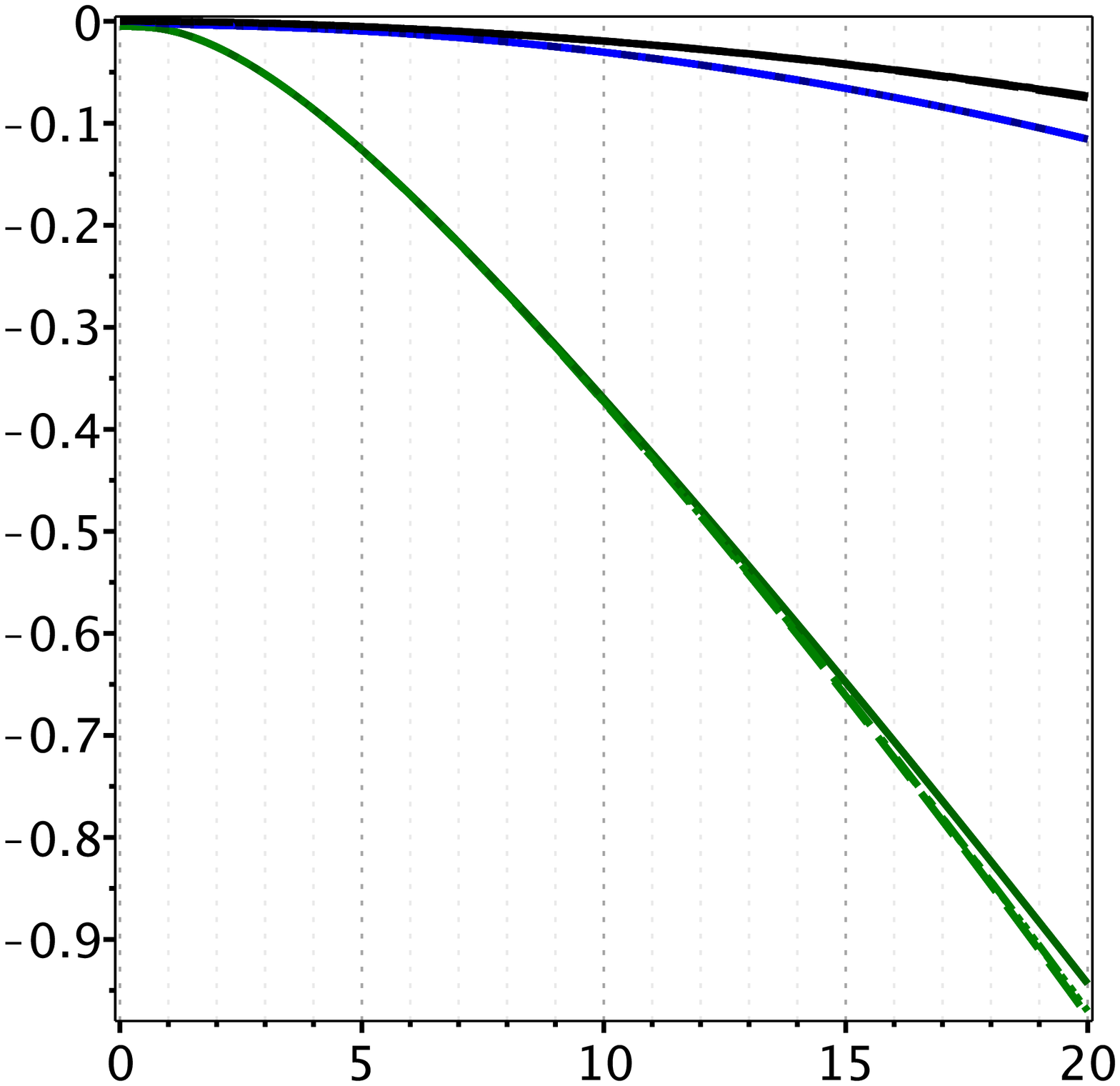}
		\put (-3,30) {\rotatebox{90}{\small Energy (eV)}}
		\put (24,2) {\small Percents of the path}
		\put (44,13) {$\Gamma \rightarrow K$}
		\put (15,85) {\large b)}
	\end{overpic}
	\hspace*{-4mm}
	\begin{overpic}[width=0.5\linewidth]{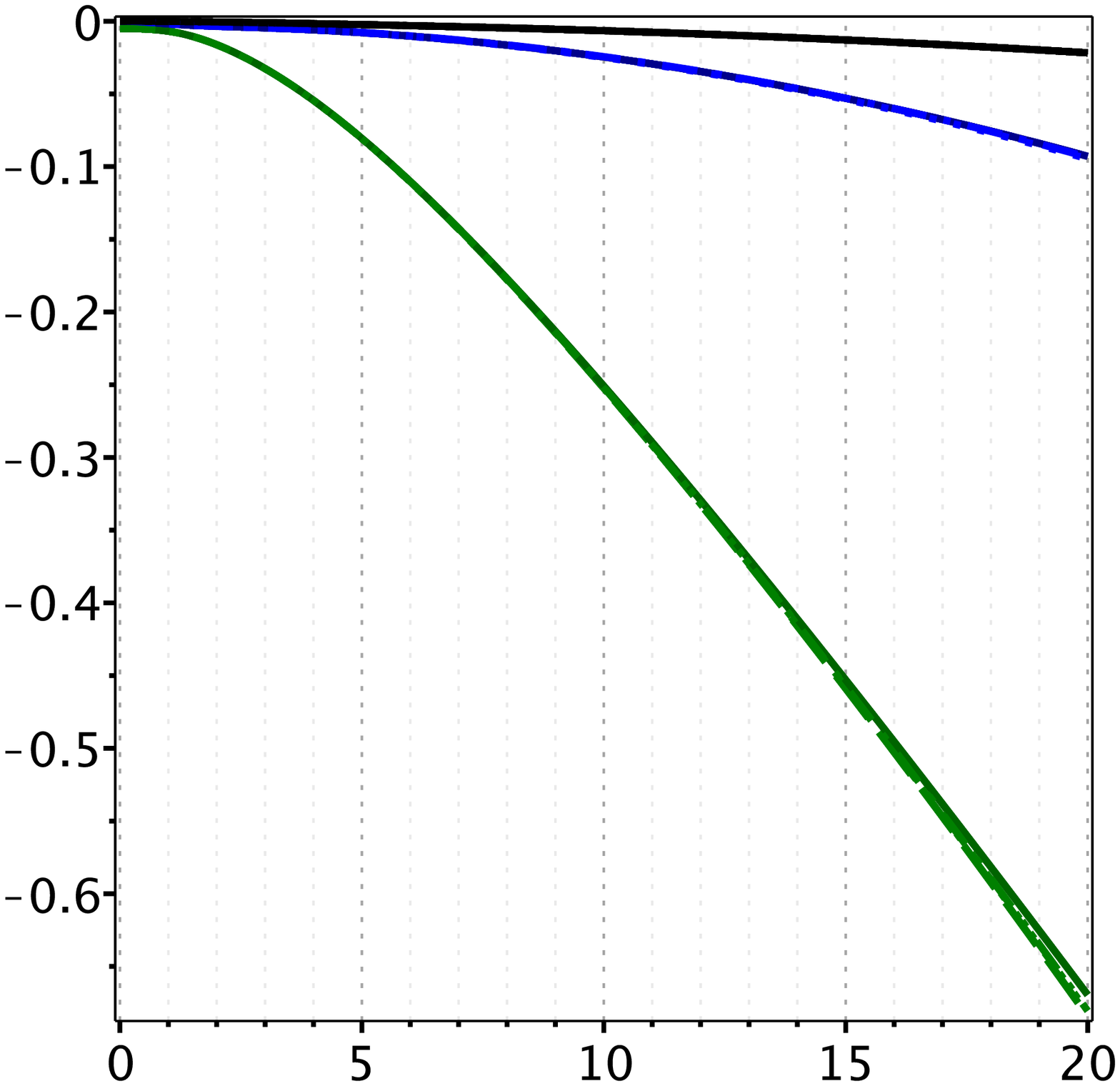}
		\put (-3,30) {\rotatebox{90}{\small Energy (eV)}}
		\put (24,2) {\small Percents of the path}
		\put (44,13) {$\Gamma \rightarrow L$}
		\put (15,85) {\large c)}
	\end{overpic}
	\hspace*{-1mm}
	\begin{overpic}[width=0.495\linewidth]{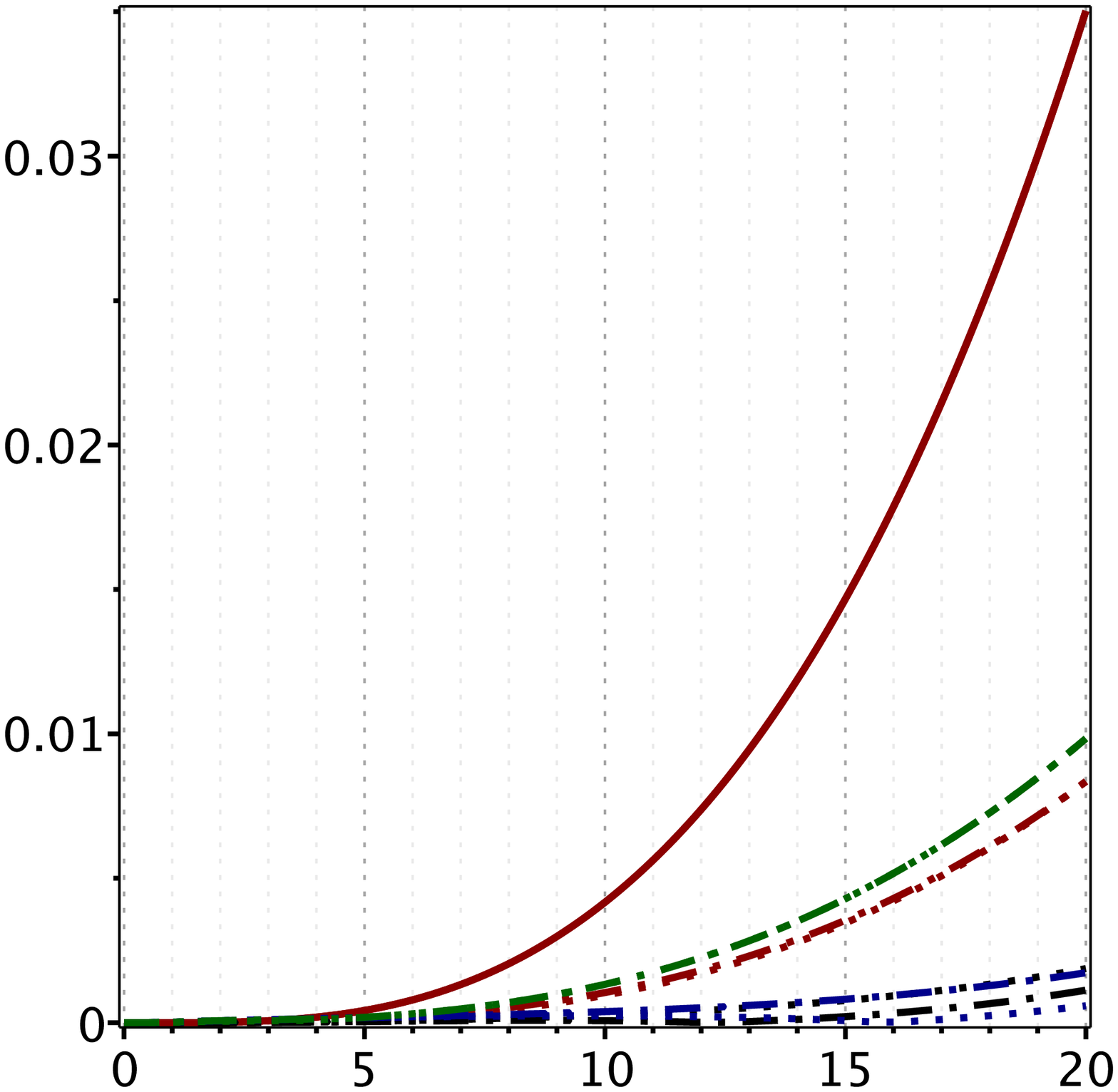}
		\put (-3,30) {\rotatebox{90}{\small Energy (eV)}}
		\put (24,2) {\small Percents of the path}
		\put (30,90) {$\Gamma \rightarrow [K,L,X]$}
		\put (15,85) {\large d)}
	\end{overpic}
	\colorbox{White}{\fbox{\parbox{22.5em}{\small%
				\centering 
				\begin{tabular}{rllll}
					Original set: & {\color{Red}\lrdashed}  CB &%
					{\color{Black}\lrdashed}  HH &%
					{\color{Blue}\lrdashed}  LH &%
					{\color{OliveGreen}\lrdashed}  SO%
					\\
					\parbox[c]{5.4em}{Rescaled set\\ (B = 0)}: &%
					{\color{Red}\lrsolid}  CB &%
					{\color{Black}\lrsolid}  HH &%
					{\color{Blue}\lrsolid}  LH &%
					{\color{OliveGreen}\lrsolid}  SO%
					\\%
					\parbox[c]{5.4em}{Rescaled set\\  (B $\approx$ 15)}: &%
					{\color{Red}\parbox[c]{12pt}{\lrdotted\\[-5pt]\lrdotdashed}}  CB &%
					{\color{Black}\parbox[c]{12pt}{\lrdotted\\[-5pt]\lrdotdashed}}  HH &%
					{\color{Blue}\parbox[c]{12pt}{\lrdotted\\[-5pt]\lrdotdashed}}  LH &%
					{\color{OliveGreen}\parbox[c]{12pt}{\lrdotted\\[-5pt]\lrdotdashed}}  SO 
				\end{tabular}		
	}}}
	\caption{Comparison of original and rescaled parameter sets for InN (color online):  original set \#\ref{mp:ZBK:GaN_LB1} (solid); rescaled set from Table \ref{tabMPZBL8x8resc} with $B=0$ (dashed) and with optimal $B=\num{15.0063671102214}$ (dotted and dash-dotted).	
		Bandstructure along the fraction of symmetry paths: a) Conduction band (CB) along $\Gamma K$; b) Valence bands along $\Gamma K$; c) Valence bands along $\Gamma L$; d) Maximum  absolute difference (direction-wise) between original and rescaled sets}
	\label{fig:ZBK_rescB_15_bs_comp:InN}
\end{figure}

The first larger value of $|B| = \num{15.0063671102214}$ for InN is a result of the error minimization over two conduction bands only.
The conduction band error err$_\text{\tiny CB} = \num{8.351977353705}$ meV signifies that the correct dispersion of CB can be recovered almost perfectly by selecting the appropriate $|B|$. 
These stated differences between the obtained dispersion and the one for original parameter sets are caused by the non-zero spin-splitting of the CB states for $B \neq 0$, which is also witnessed experimentally\cite{Zhang2008,Mei2012}. 
To illustrate the effects of spin-splitting and visualize the behaviour of adjustment error, we provide in FIG. \ref{fig:ZBK_rescB_15_bs_comp:InN} bandstructure  plots for InN along selected directions together with the plot of direction-wise maximal absolute error between the banstructure of original set \#\ref{mp:ZBK:InN_Rinke}
from Table \ref{tabMPZBL8x8} with $B=0$ and rescaled set from Table \ref{tabMPZBL8x8resc} with $B=\num{15.0063671102214}$. 

Notice from FIG. \ref{fig:ZBK_rescB_15_bs_comp:InN} b),c) that the spin-splitting is even more evident for LH and SO valence bands than for the conduction band depicted in FIG. \ref{fig:ZBK_rescB_15_bs_comp:InN} a).
Direction-wise the magnitude of CB spin-orbit splitting depends on the ratio of the individual momentum components $k_x\slash k_y$, $k_x\slash k_z$, $k_y\slash k_z$. 
It is non-zero if all these ratios are not zero, infinity or one.  

The overall eight-bands' adjustment error err$ = \num{9.839202485764}$ meV is still more three times smaller for the calculated $B$ than for $B=0$ (green dotted and dash-dotted lines vs red solid line in  FIG. \ref{fig:ZBK_rescB_15_bs_comp:InN} d) ). Now this error is dominated by the error of SO band that comes from dispersion along $\Gamma K$ direction (FIG. \ref{fig:ZBK_rescB_15_bs_comp:InN} b).  
This kind of dominance is typical for the consider materials.

\begin{figure}[t]
	\hspace*{-4mm}
	\begin{overpic}[width=0.49\linewidth]{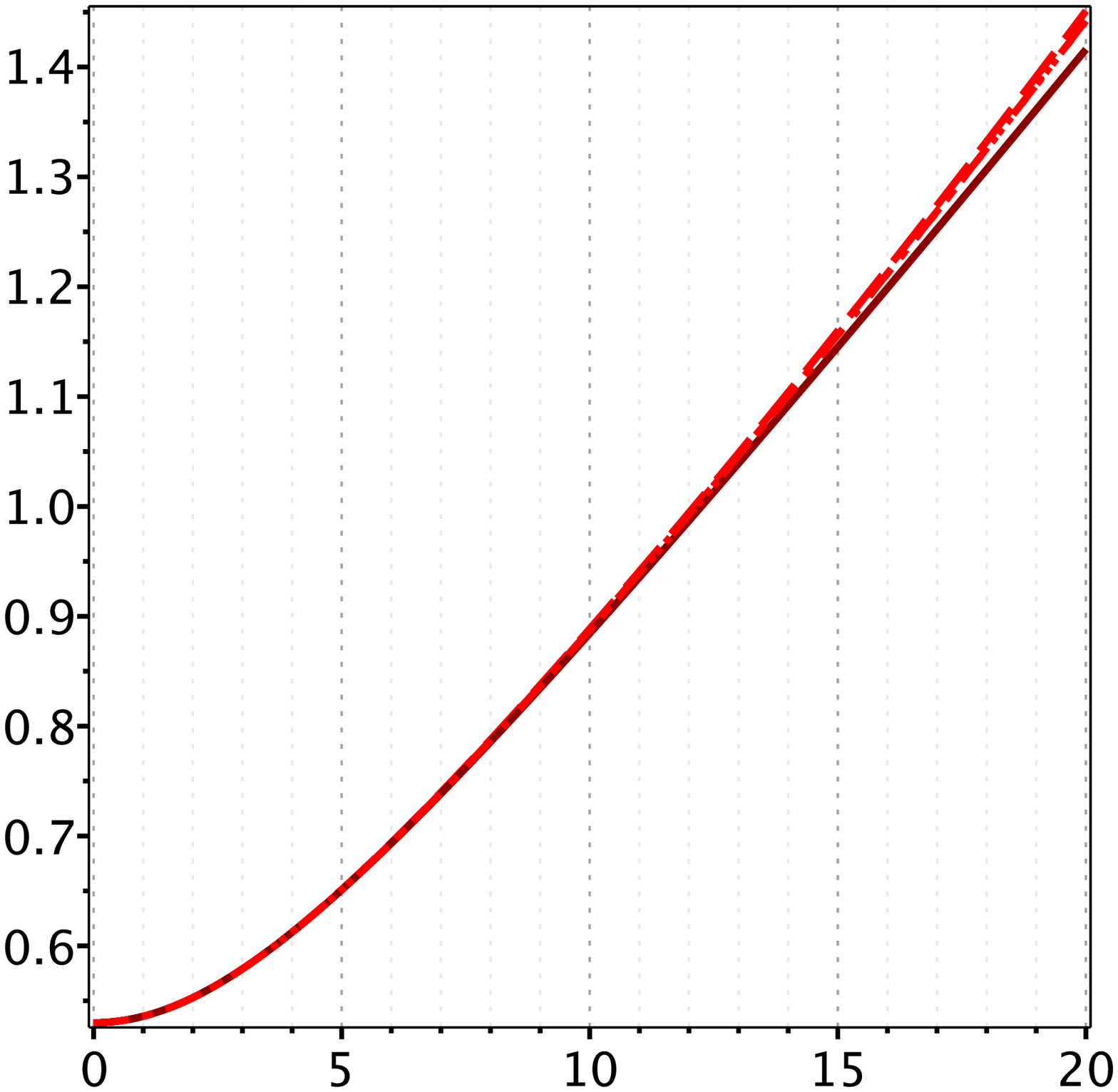}
		\put (-3,30) {\rotatebox{90}{\small Energy (eV)}}
		\put (24,2) {\small Percents of the path}
		\put (44,13) {$\Gamma \rightarrow K$}
		\put (15,85) {\large a)}
	\end{overpic}
	\hspace*{-1mm}
	\begin{overpic}[width=0.5\linewidth]{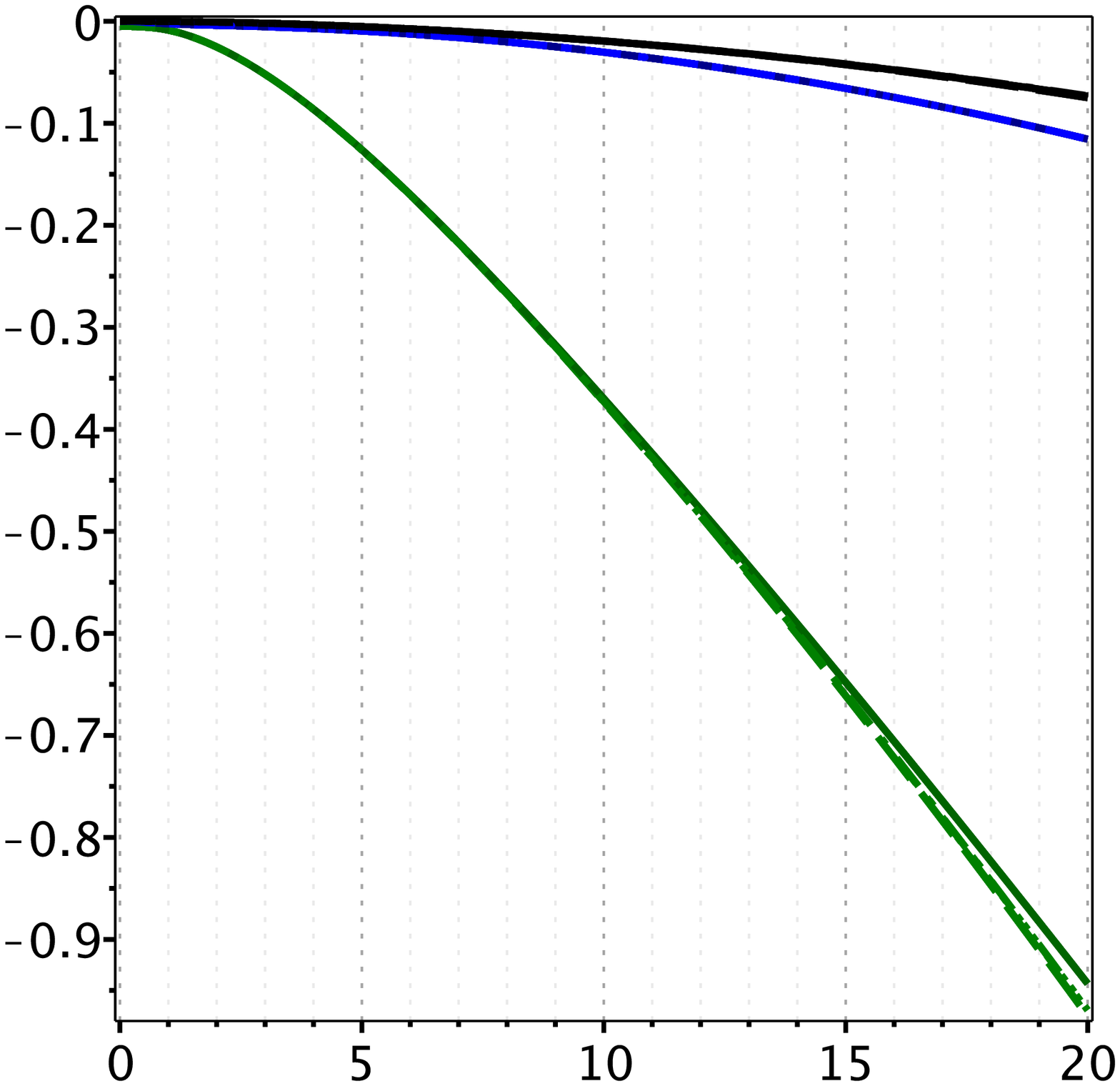}
		\put (-3,30) {\rotatebox{90}{\small Energy (eV)}}
		\put (24,2) {\small Percents of the path}
		\put (44,13) {$\Gamma \rightarrow K$}
		\put (15,85) {\large b)}
	\end{overpic}
	\hspace*{-4mm}
	\begin{overpic}[width=0.5\linewidth]{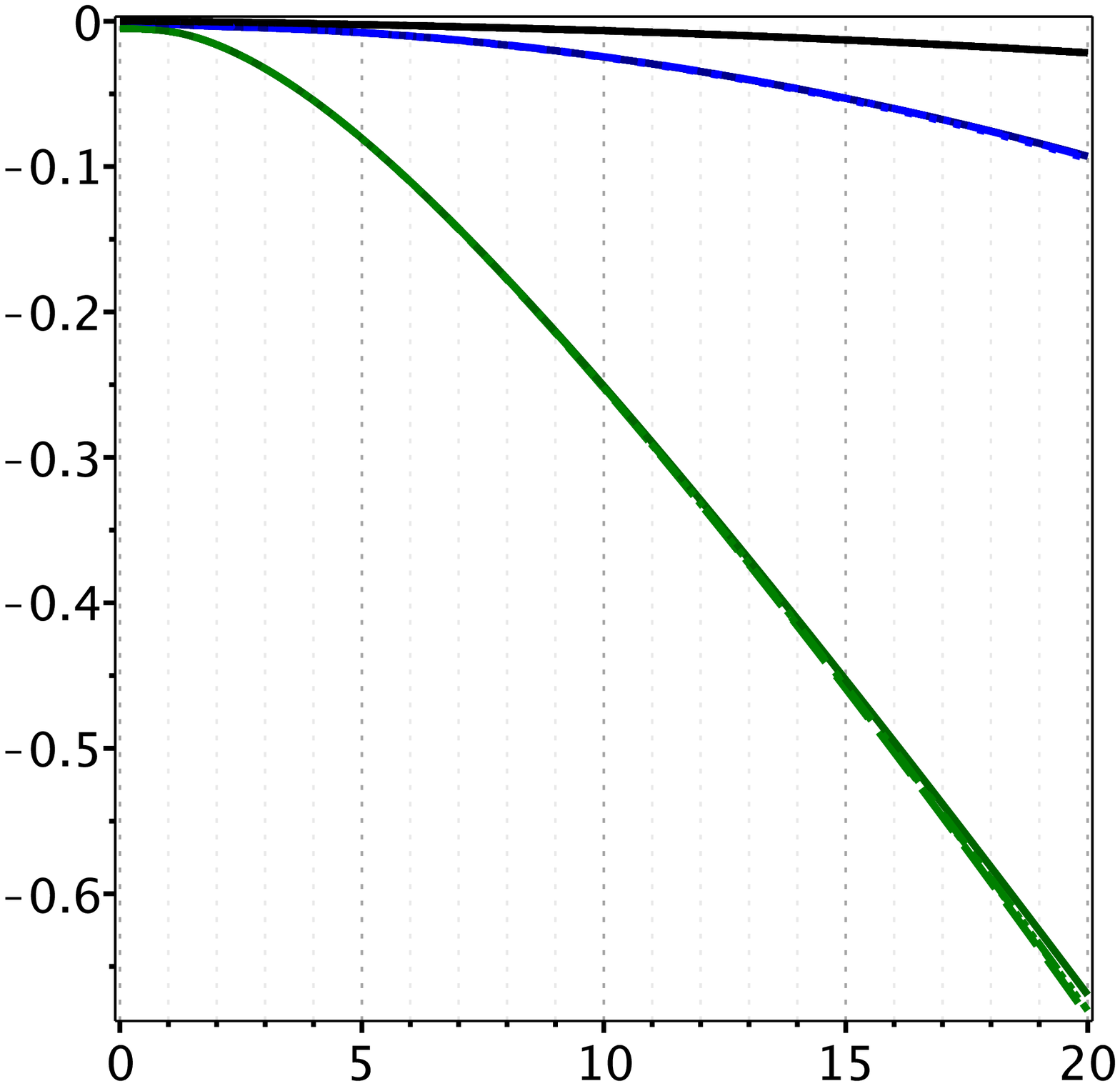}
		\put (-3,30) {\rotatebox{90}{\small Energy (eV)}}
		\put (24,2) {\small Percents of the path}
	\put (44,13) {$\Gamma \rightarrow L$}
	\put (15,85) {\large c)}
\end{overpic}
	\hspace*{-1mm}
\begin{overpic}[width=0.495\linewidth]{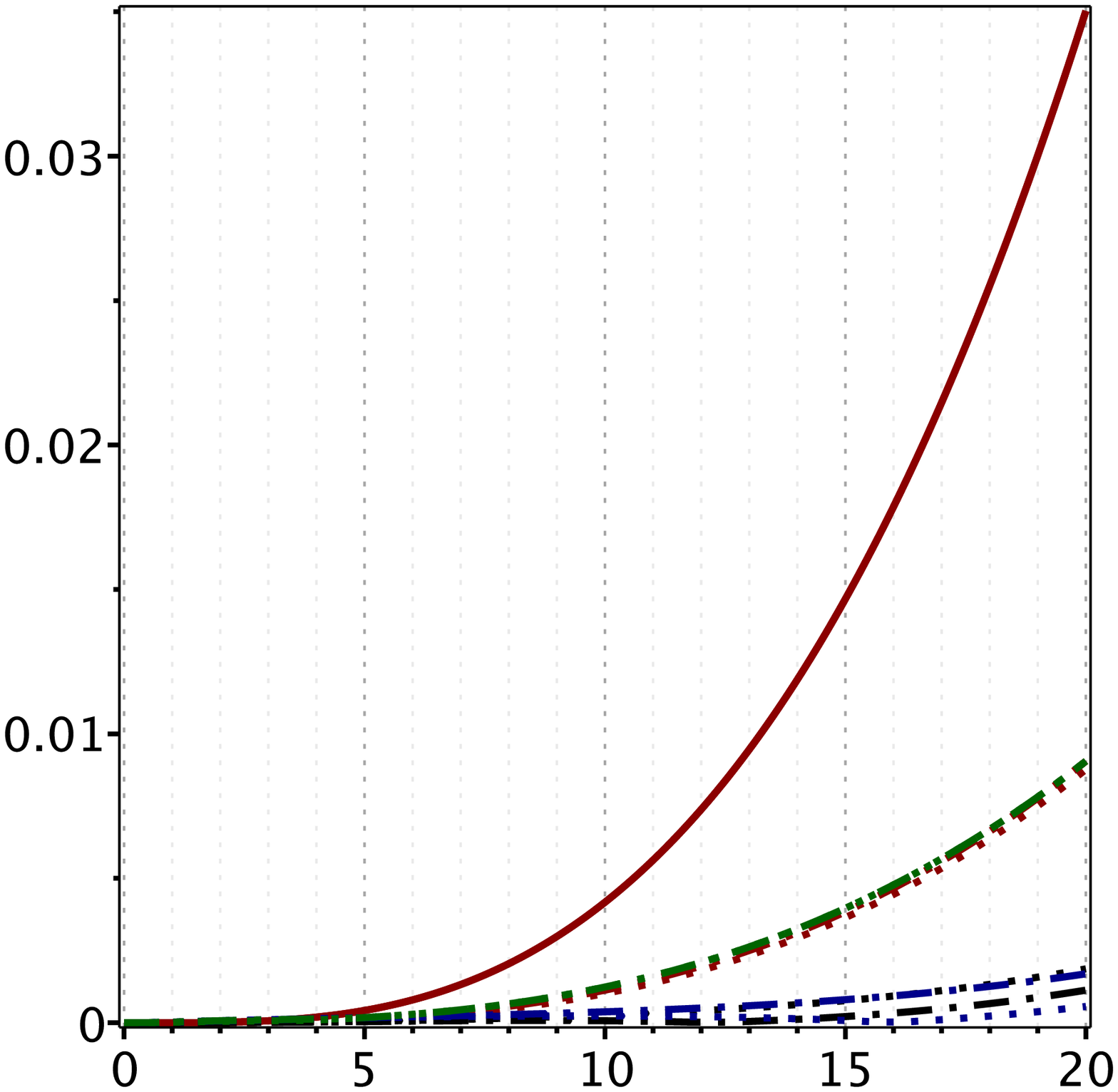}
		\put (-3,30) {\rotatebox{90}{\small Energy (eV)}}
		\put (24,2) {\small Percents of the path}
	\put (30,90) {$\Gamma \rightarrow [K,L,X]$}
	\put (15,85) {\large d)}
\end{overpic}
\colorbox{White}{\fbox{\parbox{22.5em}{\small%
			\centering 
			\begin{tabular}{rllll}
				Original set: & {\color{Red}\lrdashed}  CB &%
				{\color{Black}\lrdashed}  HH &%
				{\color{Blue}\lrdashed}  LH &%
				{\color{OliveGreen}\lrdashed}  SO%
				\\
				\parbox[c]{5.4em}{Rescaled set\\ (B = 0)}: &%
				{\color{Red}\lrsolid}  CB &%
				{\color{Black}\lrsolid}  HH &%
				{\color{Blue}\lrsolid}  LH &%
				{\color{OliveGreen}\lrsolid}  SO%
				\\%
				\parbox[c]{5.4em}{Rescaled set\\  (B $\approx$ 14.8)}: &%
				{\color{Red}\parbox[c]{12pt}{\lrdotted\\[-5pt]\lrdotdashed}}  CB &%
				{\color{Black}\parbox[c]{12pt}{\lrdotted\\[-5pt]\lrdotdashed}}  HH &%
				{\color{Blue}\parbox[c]{12pt}{\lrdotted\\[-5pt]\lrdotdashed}}  LH &%
				{\color{OliveGreen}\parbox[c]{12pt}{\lrdotted\\[-5pt]\lrdotdashed}}  SO 
			\end{tabular}		
}}}
	\caption{Comparison of original and rescaled parameter sets for InN (color online):  original set \#\ref{mp:ZBK:GaN_LB1} (solid); rescaled set from Table \ref{tabMPZBL8x8resc} with $B=0$ (dashed) and with optimal $B=\num{14.8052529908142}$ (dotted and dash-dotted).	
	Bandstructure along the fraction of symmetry paths: a) Conduction band (CB) along $\Gamma K$; b) Valence bands along $\Gamma K$; c) Valence bands along $\Gamma L$; d) Maximum  absolute difference (direction-wise) between original and rescaled sets}
	\label{fig:ZBK_rescB_14_8_bs_comp:InN}
\end{figure}

Staring from the same sets of parameters in Table \ref{tabMPZBL8x8resc}, we performed another optimization procedure with the aim of verifying at what extent the overall eight-bands' adjustment error can be minimized with help of $B$.
The resulting value of $B = \num{14.8052529908142}$ and error err$ = \num{9.05653666431}$ meV are non-significantly differ from the results of the previous optimization procedure.
The corresponding band dispersion for InN is visualized in FIG. \ref{fig:ZBK_rescB_14_8_bs_comp:InN} by using the layout of the previous figure.

The obtained minimal error is around 10 \% smaller than the  eight-bands' error of the previous optimization procedure and almost  6 times smaller than the banstructure error of rescaled parameters with zero $B$ (see the last column of Table \ref{tabMPZBL8x8resc}).

For InN, the same conclusion can be made from FIG. \ref{fig:ZBK_rescB_14_8_bs_comp:InN} d), where the band-wise error dispersion (dotted and dash-dotted lines) is plotted along with the error dispersion of the elliptic parameter set with $B=0$. 
The overall error is apparently dominated by the adjustment error of CB and SO bands (FIG. \ref{fig:ZBK_rescB_14_8_bs_comp:InN} a,b). 
The errors of other bands are below 1 meV. 

The error plots of FIG. \ref{fig:ZBK_rescB_15_bs_comp:InN} and FIG. \ref{fig:ZBK_rescB_14_8_bs_comp:InN} are also useful to quantify the magnitude of spin-spliting in CB, LH and SO bands for $B=\num{15.0063671102214}$ and $B=\num{14.8052529908142}$.
The comparison of calculated spin-splitting parameters for GaAs and AlAs from Table \ref{tabMPZBL8x8resc} and the experimental values for CB along $(1,1,0)$ direction suggests that the value of $B$ around 70-80 eV is needed for the $8 \times 8$ model to reach the reported experimental values\cite{Eppenga1988,SO_splitting_Jusserand1995,SO_splitting_Richards1996}.
For such $B$ we observed the deviation in band dispersion of around 0.35-0.5 eV from the dispersion for zero $B$.
Thus, spin-splitting errors are more dominant than the adjustment errors for selected material sets and possibly others (for the experimental values of $B$ see Table 5.5 from [\onlinecite{Cartoixa2003}] and the references wherein).
In order to achieve better accuracy with $B \neq 0$ one should use the bandstructure diagram with realistic spin-splitting of bands as an optimization target.
Having that in hands, one can possibly get better results by applying the implemented two-step adjustment procedure to other materials data entries from Table \ref{tabMPZBL8x8}  where the range $(\Delta_{05}^{\text{min}}, \Delta_{05}^{\text{max}})$ for the adjustment parameter  $\Delta_{05}$ is non-empty. 

To clarify the use of $|B|$ in the above calculations we note, that similarly to ellipticity constraints \eqref{eq:el_constrZBKane_nzB}, the sign of $B$\cite{Cardona1986} has no effect on the eigenvalues of the Hamiltonian in momentum or position representation. 
It, however, affects  the eigenstates of the Hamiltonian in both representations and therefore must be taken into consideration for experiments\cite{Chantis2008, Chantis2010} that make use of the eigenstates. 

One can further increase the accuracy of admissible parameter set of $8 \times 8$ Hamiltonian by fitting\cite{Bastos2016}  the full set of $A, \gamma_1', \gamma'_2, \gamma'_3, B$  and using inequalities \eqref{eq:el_constrZBKane_nzB} as constraints for the fitting method.
Our initial results in that direction show that this is possible for a wide range of materials. 
Besides, the adjustment with $B$ alone is not a universal substitute for the optimally fitted parameter set $A, \gamma_1', \gamma'_2, \gamma'_3, B$ because the effect of $B$ on the bandstructure disappears if two out of three momentum components are zero.

Now, let us get back to ellipticity conditions \eqref{eq:el_constrZBKane_nzB}. 
So far we used $B$ as a bandstructure fitting parameter, after the ellipticity of parameter set was established by rescaling procedure \eqref{eqConvLLm}, \eqref{eq:Kane_Ep_adj}, \eqref{eq:ZBK_delta05_sel} with zero $B$. 
The procedure with adjustment of $B$ can be used directly for the materials with the elliptic valence band part (e.g. sets \#\ref{mp:ZBK:InAs_Vurg1}, \#\ref{mp:ZBK:InAs_LB1_1}, \#\ref{mp:ZBK:AlP_Vurg1}, \#\ref{mp:ZBK:AlSb_Vurg1}). 
For such materials we can overturn the negative sign of $\lambda'_5$ and make the Hamiltonian fully elliptic by setting $B$ to the appropriate (in terms of \eqref{eq:4x4constrB}) nonzero value.
This allows to bypass the rescaling procedure altogether, which might be favourable in the light of its phenomenological nature.
Another, more physically convenient, way to increase the accuracy of $k\cdot p $ Hamiltonian is to extend its basis set by adding new energy band states. 
The resulting Hamiltonians are analyzed in the next section.

To conclude the discussion on the ellipticity of $8 \times 8$ ZB Hamiltonians we apply the direct adjustment of $B$ to several material parameter sets from Table \ref{tabMPZBL8x8}; namely sets \#\ref{mp:ZBK:InAs_Vurg1}, \#\ref{mp:ZBK:InAs_LB1_1} for  InAs, set \#\ref{mp:ZBK:AlP_Vurg1} for AlP and set \#\ref{mp:ZBK:AlSb_Vurg1} for AlSb. 
Only those listed parameter sets yield elliptic valence band part of the Hamiltonian, i.e. the corresponding distance $d=0$ in Table \ref{tabMPZBL8x8}. 
Notably, for each of the three materials the first listed set was reported in  [\onlinecite{Vurgaftman2001}] -- the work that is  highly regarded as a source of overall physically consistent material parameters.
For InAs,  the direct adjustment of $B$ is the only option to obtain the admissible parameter sets based on the data from Table \ref{tabMPZBL8x8}, because the admissible range $(\Delta_{05}^{\text{min}}, \Delta_{05}^{\text{max}}) \ni \Delta_{05}$ is empty for all its four table entries.
For each above mentioned parameter sets we performed bandstructure error minimization procedure over the interval  $|B| \in (|B|_\text{\tiny min}, \infty)$ and reported the resulting value $|B|$, along with the errors in Table \ref{tabMPZBL8x8optBd}. Here $|B|_\text{\tiny min}$ is the minimal solution of \eqref{eq:4x4constrB}, with $0.1$  subsituted in place of $0$ in the right-hand side to accommodate for possible numerical errors.
\begin{table}[ht]
	\centering
	\caption{Results of direct bandstructure error minimization procedures based on the adjustment of $B$.}\label{tabMPZBL8x8optBd}   
	\tabcolsep=0.5em 
	\sisetup{
		round-mode      = places,
		round-precision = 3,
		table-number-alignment = left,
		table-figures-integer  = 1,
		table-figures-decimal  = 2,
		table-align-text-post = false,
		table-sign-mantissa = false,
		tight-spacing = true,
	}
	\begin{threeparttable}
	\csvloop
		{
			file=ZB_8x8_B_direct_errors.csv,
			no head, column count=17, column names reset,
			column names={1=\one, 2=\two, 3=\three, 4=\four,5=\five, 6=\six, 7=\seven, 8=\eight,9=\nine, 10=\ten,11=\eleven,12=\twelve
			},
			tabular={@{\hskip 1pt}l@{\hskip 10pt}|@{\hskip 6pt}S[round-precision = 2, table-format=2.2]S[round-precision = 2, table-format=2.2]SSSSSS@{\hskip 0pt}},
			table head=\toprule[0.3pt]\midrule[0.3pt]\bottomrule %
			El\tnote{a} & {$|B|_\text{\tiny min}$} & {$|B|$\tnote{b}} %
			
			& {err\tnote{c}} & {err$_\text{\tiny CB}$\tnote{d}} & {err$_\text{\tiny HH}$\tnote{d}} & {err$_\text{\tiny LH}$\tnote{d}} & {err$_\text{\tiny SO}$\tnote{d}}%
			\\\hline\midrule\bottomrule, 
			command= \two\tnote{\ref{\twelve}} %
			 & \three & \six & \seven & \eight & \nine & \ten & \eleven, 
			table foot=\toprule\midrule[0.3pt]\bottomrule[0.3pt],
			filter = \(\thecsvinputline>1\) \and \(\thecsvinputline<53\)%
		}
	\begin{tablenotes}\footnotesize
			\item [a] Refer to the original dataset number from Table \ref{tabMPZBL8x8}
			\item [b] $B$ that minimizes the banstructure error
			\item [c] The minimal value of the error in eV calculated for the eight bands with a given $B$ over 20 \% of the paths $ \Gamma L$, ${ \Gamma K}$ and ${ \Gamma X}$
			\item [d] The errors {err$_\text{\tiny CB}$}, err$_\text{\tiny HH}$, {err$_\text{\tiny LH}$}, {err$_\text{\tiny SO}$} (all in eV) of conduction bands, heavy holes, light holes and split-off bands accordingly
		\end{tablenotes}
	\end{threeparttable}
\end{table}

For brevity we do not provide bandstructure or error plots based on the data from Table \ref{tabMPZBL8x8optBd}. 
Alternatively, we supplied the errors of CB, HH, LH, SO bands as separate entries in the table. 
For InAs and AlP the introduced bandstructure error is dominated by the differences in CB, SO and LH bands. 
The situation is different for AlSb for which the main contribution to the error comes from LH bands.

The results reported in Table \ref{tabMPZBL8x8optBd} clearly indicate that set \#\ref{mp:ZBK:AlP_Vurg1} for AlP along with $B=\num{5.272372155}$ lead to the smaller bandstructure adjustment error than the same-material set reported in Table \ref{tabMPZBL8x8resc} with $B=0$.
We recommend this for simulations based on the $8 \times 8$ ZB Hamiltonian\cite{kp8ProperKane_Bahder90} in the position representation.
For InAs we suggest using the set \ref{mp:ZBK:InAs_Vurg1} with $B=\num{25.89773358}$.
The set obtained as a result of two step optimization, reported above, is the most optimal for InN. 
For the rest of materials analyzed in this work we recommend the sets from Table \ref{tabMPZBL8x8resc}.

\section{Ellipticity of $14 \times 14$ band models}\label{sec14x14}
In this section we focus our attention on the ellipticity of two $14 \times 14$ ZB Hamiltonians that are frequently used in the literature\cite{14x14Braun1985,kp14Zawadzki1985,kp14Pfeffer90,kp14Mayer91,kp14Zawadzki1992,kp14Pfeffer96,kp14Pfeffer2006,kp14Cavassilas01,kp14Bhat2005,kp14Pfeffer2006,kp14kp6Fishman03,kp14kp8Gladysiewicz2015}.
These two models are based on the extended basis set: six $p$-like valence band states and two $s$-like conduction states comprising the basis  of $8 \times 8$ Hamiltonian studied in the previous section, plus six additional $p$-like conduction band states. 
They are introduced to better describe anisotropy of conduction band in the materials like GaAs, InP, InSb, where it is evidenced experimentally\cite{14x14Braun1985,kp14Mayer91,kp14Zawadzki1992,kp14Pfeffer96}.

The first Hamiltonian proposed by  W. Zawadzki, P. Pfeffer and H. Sigg in  [\onlinecite{kp14Zawadzki1985}] and then extended\cite{kp14Zawadzki1992,kp14Pfeffer96} to account for the influence of the out-of-basis bands perturbatively.  
We base our analysis on this later extended version described by equation (5) from [\onlinecite{kp14Pfeffer96}]. 
The calculated\cite{Note1} eigenvalues $\lambda''_1 - \lambda''_5$ of the  quadratic form associated with this $14 \times 14$ ZB Hamitonian are as follows 
\begin{equation}\label{eigZB14P}
\begin{array}{rl}
\lambda''_1 &= E(-\gamma''_1 - 4\gamma''_2 - 6\gamma''_3)\\[2pt]
\lambda''_2 &= E(-\gamma''_1 - 4\gamma''_2 + 3\gamma''_3)\\[2pt]
\lambda''_3 &= E(-\gamma''_1 + 2\gamma''_2 + 3\gamma''_3)\\[2pt]
\lambda''_4 &= E(-\gamma''_1 + 2\gamma''_2 - 3\gamma''_3)\\[2pt]
\lambda''_5 &= E. 
\end{array}
\end{equation}

The CB part of the Hamiltonian is elliptic by design, since $\lambda''_5 >0$ independently of materials parameters.  
The ellipticity of the valence-band part is guaranteed when $\lambda''_1 - \lambda''_4$ are all negative simultaneously.
So in the end, we are getting exactly the same ellipticity conditions as for the $6 \times 6$ ZB Hamiltonian, albeit with the different Luttinger-like parameters (compare the above $\lambda''_1 - \lambda''_4$ with $\lambda'_1 - \lambda'_4$ from \eqref{eig_LK}).
These new Luttinger-like parameters $\gamma''_1, \gamma''_2, \gamma''_3$ can be obtained from the conventional Luttinger parameters by subtracting from $\gamma_1, \gamma_2, \gamma_3$ the contributions of $p$-like CB bands, that are no longer treated perturbatively. 
More precisely,
\begin{equation}\label{eqConvLLpp}
\begin{split}
\gamma''_1 &= \gamma'_1 - \frac{Q^2}{3 E E'_0} - \frac{Q^2}{3E(E'_0 + \Delta'_0)},\\
\gamma''_2 &= \gamma'_2 + \frac{Q^2}{6 E E'_0},\quad 
\gamma''_3 =\gamma'_3  - \frac{Q^2}{6 E E'_0}.
\end{split}
\end{equation}
Here 
$E_0$ is a fundamental bandgap, $E'_0$ is a gap between first two bottommost conduction bands, $\Delta_0$, $\Delta'_0$ are the  spin-splitting parameters of the valence and conduction bands correspondingly,  $Q$ is the interband momentum matrix element (see Fig 1 in \onlinecite{kp14Pfeffer96}). 
For a complete description of the Hamiltonian one additionally needs to define other material dependent parameters $P'_0, \ovl{\Delta}, \kappa, C_k$. 
Those are determined by fitting the bandstructure to experimental data\cite{kp14Pfeffer96}.
For that reason, we are focused only on the sources where the full sets of fitted Hamiltonan parameters have been reported in the context of the considered $14 \times 14$ model.

Beside the sets $\gamma''_1, \gamma''_2, \gamma''_3$ from the original paper\cite{kp14Pfeffer96}, that provides them for GaAs and InP explicitly, we used the parameters sets from J.-M. Jancu et al. [\onlinecite{kp14Jancu2005}] and recommended there Luttinger parameters from [\onlinecite{Vurgaftman2001}] to calculate the respective values of $\gamma''_1, \gamma''_2, \gamma''_3$ for GaAs, AlAs, InAs, GaP, AlP, InP, GaSb, AlSb and InSb.
All calculated parameters along with the results of ellipticity analysis are collected in Table  \ref{tabMPZBL14x14resc}.
\begin{table}
	\centering \caption{The material data for $14 \times 14$ ZB Hamiltonian\cite{kp14Pfeffer96}, $d$ -- distance from
			the point $(\gamma''_{1},\gamma''_{2},\gamma''_{3})$ to the feasibility region $\Lambda_{-}$. 
			The positive values of $\lambda''_1/E, \lambda''_2/E, \lambda''_3/E, \lambda''_4/E$ are printed in bold.}
	\label{tabMPZBL14x14resc} 
	\tabcolsep=0.5em 
	\newcounter{zbptabno}
	\begin{threeparttable}
		\csvloop
		{
			file=ZB_14x14_B_0_selected.csv,
			no head, column count=17, column names reset,
			before reading=\centering\sisetup{table-number-alignment=center, detect-weight=true, detect-inline-weight = text, round-precision = 2, table-figures-integer  = 1, table-figures-decimal  = 2},
			column names={1=\one, 2=\two, 3=\three, 4=\four,5=\five, 6=\six, 7=\seven, 8=\eight,9=\nine, 10=\ten,11=\eleven,12=\twelve, 13=\thirteen, 14=\fourteen, 15=\fifteen		
			},
			before line={\refstepcounter{zbptabno}},
			tabular={@{\hskip 0pt}r@{\hskip 3pt}l|@{\hskip 8pt}SSSS@{\hskip 5pt}S@{\hskip 4pt}S@{\hskip 4pt}S@{\hskip 3pt}S[table-format=1.2]@{\hskip 1pt}},
			table head=\toprule[0.3pt]\midrule[0.3pt]\bottomrule %
			\# & El\tnote{a} %
			& {$\gamma''_1$} & {$\gamma''_2$} & {$\gamma''_3$} %
		    & {$\lambda''_1/E$} %
			& {$\lambda''_2/E$} %
		    & {$\lambda''_3/E$} & {$\lambda''_4/E$} %
		    & {$d$} 
			\\\hline\midrule\bottomrule, 
			command= {\thezbptabno\fifteen} & {\two\thirteen} %
			& \three & \four & \five %
			& \ifdimgreater{0 mm}{\six mm}{\six}{\bfseries \six} %
			& \ifdimgreater{0 mm}{\seven mm}{\seven}{\bfseries \seven} %
			& \ifdimgreater{0 mm}{\eight mm}{\eight}{\bfseries \eight} %
			& \ifdimgreater{0 mm}{\nine mm}{\nine}{\bfseries \nine} %
			& \ten, 
			table foot=\toprule\midrule[0.3pt]\bottomrule[0.3pt],
			filter = \(\thecsvinputline>1\) \and \(\thecsvinputline<53\)%
		}
		\begin{tablenotes}\footnotesize
			\item [a] Set 1 from [\onlinecite{kp14Pfeffer96}] ($\alpha  = 0.065$)
			\item [b] Set 2 from [\onlinecite{kp14Pfeffer96}] ($\alpha  = 0.085$)
			\item [c] Parameters obtained via \eqref{eqConvLLpp} from the data in [\onlinecite{kp14Jancu2005}] and [\onlinecite{Vurgaftman2001}]			
			\item [d] Set 1 from [\onlinecite{kp14Pfeffer96}] ($\alpha  = 0.12$)
			\item [e] Set 2 from [\onlinecite{kp14Pfeffer96}] ($\alpha  = 0.2$)
		\end{tablenotes}
	\end{threeparttable}
\end{table}

Each of the considered in Table \ref{tabMPZBL14x14resc} sets fails two out of four ellipticity constraints except the sets for AlAs, AlP and InSb, for which three ellipticity constrains are violated.
For GaAs set \ref{mp:ZB14:GaAs1_Pfeffer96} from Table \ref{tabMPZBL14x14resc} we can confirm a reduction of distance $d$ to the feasibility region $\Lambda_{-}$ in the space $\gamma''_1, \gamma''_2, \gamma''_3$  compared to the best of GaAs  sets for $6\times 6$ and $8 \times 8$ Hamiltonians.
This set and set \#\ref{mp:ZB14:InP1_Pfeffer96} are taken from the original work. 
Both sets were calculated using cyclotron resonance experiments\cite{kp14Pfeffer96}.
The second pair of sets \#\ref{mp:ZB14:GaAs2_Pfeffer96}, \#\ref{mp:ZB14:InP2_Pfeffer96}, which are deemed more consistent experimentally\cite{kp14Gorczyca1991}, is slightly off the region $\Lambda_{-}$; but the corresponding values of $d$ are within the range of the same-material values of $d$ from Table \ref{tabMPZBL8x8}.
Parameter sets for other materials are even further away from $\Lambda_{-}$ than the un-rescaled same-materials  sets for $8 \times 8$ Hamiltonian. 

The observed increase of the distance to $\Lambda_{-}$  seems to be theoretically unfounded, especially in the view of formula \eqref{eq_PO}. 
Recall, that the relative norm\cite{Kato1} of the perturbative term from \eqref{eq_PO} should decrease after eigenstates are moved from perturbative class (class B) in to the basis (class A). 
It can be explained as follows. 

In the $8 \times 8$ model the influence of valence bands on the CB states was represented directly by the parameters $P_0$, $B$ and, we suppose, indirectly by the perturbative CB parameter $A'$.
The absence of $A'$ in the CB eigenvalue from \eqref{eigZB14P} suggest that the $14 \times 14$ model was derived under assumption that $A'$ depends on the upper CB states only, now included in the basis.
In such a situation, all cross-influence between valence and conduction bands are incorporated into $P$ and $Q$ by using fitting to experimental data.
Then it is propagated to $\gamma''_1, \gamma''_2, \gamma''_3$ with help of formula \eqref{eqConvLLpp}.
But the terms $\gamma'_1, \gamma'_2, \gamma'_3$ in the right-hand side of \eqref{eqConvLLpp} were fitted to experiments under assumption of the non-zero valence band contribution to $A'$.
That explains why the parameter triplets $\gamma''_1, \gamma''_2, \gamma''_3$ for materials with smaller fundamental bandgap $E_0$ (InAs, GaSb, InSb) end up having larger $d$.

On the other hand, the conduction band states in the materials with larger $E_0$ (AlAs, AlP, AlSb) may in reality be influenced by the higher bands not included in the basis.
That influence is assumed to be zero in the model, because CB eigenvalues are equal to $E$ even for the newly included in the basis $p$-like bands.
If non-negligible, the influence is accounted by $P,Q$ and then propagated to $\gamma''_1, \gamma''_2, \gamma''_3$ by the mechanism described above. 
That explain the increase in $d$ for large-bandgap materials from Table \ref{tabMPZBL14x14resc} (AlAs, AlP, AlSb).

For some parameter sets from  Table \ref{tabMPZBL14x14resc} the ellipticity might be corrected by rescaling of $P,Q$ in a way similar to the rescaling procedure from Section \ref{sec8x8}.
This will, of course, affect the accuracy of bandstructure and therefore must involve the optimization procedure with respect to the parameters $P,Q$ and possibly $\gamma''_1, \gamma''_2, \gamma''_3$, if our hypothesis holds true. 

Now we proceed to the second implementation of $14 \times 14$ ZB Hamiltonian model.
The initial version of this model was derived by U. R\"ossler using the theory of invariants\cite{kp14Roessler1984} and then extended in the work of H. Mayer and U. R\"ossler\cite{kp14Mayer91} by adding first-order perturbative corrections to the lowest conduction and upper valence bands. 
The most recent version of the Hamiltonian was provided by R. Winkler\cite{kp8x84x46x6_sp_bc_Winkler2003}. 
It additionally includes the second order conduction-valence band mixing parameters similar to $B$ from Kane Hamiltonian \eqref{eqHKane4}.

All three mentioned versions of $14 \times 14$ ZB Hamiltonian are connected by the common assumption that six second-order diagonal terms related to the newly added $p$-like CB states are neutralized by the counter-influence of other bands,  e.g. the representation of $H_{8c8c}$, $H_{8c8c}$ from Table C.5 of [\onlinecite{kp8x84x46x6_sp_bc_Winkler2003}]. 
In terms of ellipticity such an assumption results in the presence of zero eigenvalue among the set of eigenvalues of the  quadratic form associated with this implementation. 
Our calculations\cite{Note1} confirm that. 
Therefore, this Hamiltonian is not elliptic by design. 

It is worthwhile pointing that, unlike first, the second implementation of $14 \times 14$ Hamiltonian\cite{kp14Roessler1984,kp14Mayer91,kp8x84x46x6_sp_bc_Winkler2003} can be regarded as an extension of Kane model studied in section \ref{sec8x8}. 
The Hamiltonian contains the perturbative correction to $s$-like CB states and the conduction-valence band mixing parameters. So, all inter-band interaction effects embodied in the $8 \times 8$ representation\cite{kp8ProperKane_Bahder90} can be properly accounted for. 
In our opinion, two analyzed implementations of the $14 \times 14$  Hamiltonian model are less universal material-wise than $8 \times 8$ Hamiltonians, despite being more accurate at describing CB related phenomena\cite{kp14kp6Fishman03,kp14kp8Gladysiewicz2015}.
For such higher band models, the assumptions regarding the interactions of in-basis conduction band states require a revision. 

\section*{Conclusions}
We performed a systematic study of ellipticity conditions for $6 \times 6$, $8 \times 8$, $14 \times 14$ $k \cdot p$ Hamiltonians in the bulk zinc blende crystals.
The conditions take roots in the fundamental axioms of quantum mechanics concerning the description of observable states and properties of Hamiltonian as a differential operator. 
They appear in the form of constraints on the values of material parameters pertaining to the second-order-in-$k$ terms from the Hamiltonian in the momentum representation.  

For $6 \times 6$ and $8 \times 8$ models we examined an extensive number of parameter sets for GaAs, AlAs, InAs, GaP, AlP, InP, GaSb, AlSb, InSb, GaN, AlN, InN and C that are gathered from the widely accepted  sources of reference literature on material parameters\cite{LB1,Vurgaftman2001,Madelung2004}  
The results of the performed analysis reaffirm earlier conclusions on the violation of Hamiltonian ellipticity\cite{Veprek07,Veprek09} and its cause\cite{kpEllptArxSytnyk2010,MThSytnyk2010}.
Furthermore, we demonstrated that this violation is a much more common problem material-wise. 
Among all analyzed materials only carbon has parameter sets that make $6 \times 6$ Hamiltonians elliptic and therefore admissible from a theoretical point of view. 
Other sets of material parameters incur violation of one out of four ellipticity constraints: $2\gamma_2-\gamma_1  + 3\gamma_3 < 0$.
This can be traced to a non-negligible influence of conduction bands on the heavy-hole  and light-hole, accounted perturbatively in $\gamma_1,\gamma_2,\gamma_3$. 
We conclude that this model is not accurate enough to describe all considered bands reliably and to remain elliptic at the same time. 

The situation becomes more complex for $8 \times 8$ Hamiltonians, where the bottom-most conduction band is included into the basis.
None of the analyzed parameter sets are admissible, because the conduction-band ellipticity constraint is violated for all sets, when the absence of inversion asymmetry ($B=0$) is assumed.
However, the degree of non-ellipticity in the valence-band part of the Hamiltonian, which we characterize in terms of distance to the feasibility region $\Lambda_{-}$ in the space of Luttinger-like parameters,  decreases. 
Several parameter sets for InAs, AlP and AlSb satisfy the ellipticity constraints for the valence-band part. 
It corroborates the evidence on the perturbative source of non-ellipticity. 
We note, that in the case of $B=0$, these constraints have the same structure as the ellipticity constraints for $6 \times 6$ Hamiltonians.

As one possible way to remedy the situation with the lack of ellipticity in the $8 \times 8$ model we propose a parameter rescaling procedure. 
It is based on the idea of adjusting the first-order conduction-valence mixing parameter $P_0$ to change $A, \gamma'_1,\gamma'_2,\gamma'_3$ and make the Hamiltonian elliptic. 
The proposed here rescaling procedure accounts for a full set of ellipticity constraints and thus improve the previous approaches\cite{Foreman_sp_97,Birner2014}, targeted solely at imposing the conduction band constraint $1+A>0$. 

The results of the rescaling procedure for all materials except InAs, are presented in Table \ref{tabMPZBL8x8resc}.
Each of the admissible sets is made via \eqref{eq:ZBK_delta05_sel} from one of the original sets that lead to a minimal absolute difference in the parameter $A$ per material. 
Nothwithstanding the attempt to minimize the effects of rescaling on the bandstructure, these effects are negligible ($\leq 11$ meV) only for AlP, AlSb, AlAs, InP and GaN (see FIG. \ref{fig:ZBK_resc_bs_comp:GaN}). 
They may be considered small ($\leq 50$ meV) for GaP, InSb, InN and can not be ignored for the rest of materials from the table (see FIG. \ref{fig:ZBK_resc_bs_comp} for visual comparisons).
For them, the rescaling leads to a noticeable change in the conduction band dispersion. 
Heavy hole~(HH) and light hole~(LH) bands remain visually unaffected even though the differences are non-zero. 
The rescaling also causes an increase in the curvature of the split-off~(SO) band which makes it the main source of total valence-band adjustment error. 
For all mentioned materials excluding GaP and AlN, the magnitude of this error is proportional to the relative change in $E_p$.
Therefore, in most cases the Hamiltonian based on the new parameters is elliptic, yet incapable of reliably  describing the conduction-valence band transition phenomena, except those occurring at the band edge.

In attempt to counter for the observed bandstructure discrepancies and to derive the admissible parameter set for InAs  we consider the use of $B$ as an additional adjustment parameter.
This requires a generalization of the ellipticity conditions for the $8 \times 8$ Hamiltonian to the case of non-zero $B$.
Recall, that, due to the inversion asymmetry, this case is theoretically more relevant to the majority of zinc blende materials. 
The form of the generalized ellipticity conditions allows us to draw two important conclusions. 

First, setting $B$ to some nonzero value will not break ellipticity of $8 \times 8$ Hamiltonians if the parameter set -- the Hamiltonian is based upon -- is admissible with zero $B$.
We use this property to calculate two distinct values of $B$ for the materials. 
The larger value of $B$ minimizes the error of conduction band and makes the errors in the dispersion of other (most notably SO) bands larger. 
We left aside a discussion on physical relevance of the calculated values of $B$ and presented the bandstructure plots with $B\neq 0$ for InN only. 
Our intent here has been to show that $B$-adjustment can be used to partially correct the bandstructure distorted by rescaling.


Second, the parameter $B$ could not be set to zero for the materials where $1+A'<0$ without sacrificing ellipticity of the $8 \times 8$ Hamiltonian.
At the same time, the adjustment of $B$ can be used to correct the ellipticity of CB part of the Hamiltonian provided that the valence-band ellipticity constaints are fulfilled by $\gamma'_1,\gamma'_2,\gamma'_3$.
We discovered four parameter sets for InAs, AlP, AlSb where this is true.  
These parameter sets are highly regarded as overall physically consistent\cite{Vurgaftman2001}.
Four minimally admissible values of $B$ that complement each of the mentioned sets to  make the Hamiltonian elliptic, are collected in Table \ref{tabMPZBL8x8optBd}.

Besides being the source of admissible $B$, the data from Table \ref{tabMPZBL8x8optBd} illustrates that the admissible sets obtained by the rescaling procedure are not the best option in terms of the bandstructure fit, at least for AlP.
We postulate that
there exist admissible parameters of $8 \times 8$ Hamiltonian\cite{kp8ProperKane_Bahder90} better in terms of bandstrcuture error for other materials too.
To find them one should fit the Hamiltonian bands' dispersions to the spin-resolved bandstructure by adjusting the entire set of the Hamiltonians parameters simultaneously and by using the ellipticity conditions as constraints for the fit.
This idea is supplemented by the fact, that the ellipticity region in the space of parameters that satisfy the constraints is convex and connected.
The results of such fitting procedure will be also useful to quantify the limits of this and other $k \cdot p$ models, assuming  that the bandstructure used as a fitting target is reliable\cite{bsexpOkuda2017}.
This would constitute an important for the future studies.

Finally, we analyzed two popular implementations of the $14 \times 14$ ZB Hamiltonian model\cite{kp14Pfeffer96,kp8x84x46x6_sp_bc_Winkler2003}.
The ellipticity of the first implementation is described by precisely the same set of constraints as for the $6 \times 6$ model, but written in terms of the reduced Luttinger-like parameters $\gamma''_1,\gamma''_2,\gamma''_3$. 
Unfortunately, the parameter analysis shows that none of the available sets for a studied list of materials is admissible in terms of the Hamiltonian ellipticity.
We conclude that an overly-strict set of assumptions regarding the perturbative influence of outer bands on the model's conduction bands is to be responsible for the lack of ellipticity.  
The second analysed $14 \times 14$ Hamiltonian is more general in that regard.
It is however non-elliptic by design owing to the fact that the second-order-in-k terms are zero for three upper $p$-like conduction bands.
    
Based on the supplied evidence we surmise that both $14 \times 14$ implementations are less universal than the previously studied $8 \times 8$ Hamiltonian.
The revision of indicated assumptions and, perhaps, some unifications are necessary to bring these extended models 
to a strict theoretical ground. 

The analysis conducted in this paper covers possible extensions of the considered models, such as the inclussion strain-stress, electromagnetic or other phenomena, as long as such extensions do not change the structure of second-order-in-$k$ terms of the Hamiltonian.
The analysis can be easily transferred to the cases when momentum quantization is applied in one- or two- dimensions (quantum wells, wires, etc). 
It can also be applied to the materials, that are intrinsically non-three dimensional (non-3D), like graphene, silicene or others;  especially given that many high-accuracy bandstructure diagrams for those kind of materials are readily available\cite{Massatt2018,Pan2014,Lin2018}.

We note that ellipticity conditions stated here are not valid for anything other than the considered three dimensional $k\cdot p$ Hamiltonians for ZB crystals.
The whole analysis will have to be repeated in each specific non-3D case.

The applications to materials with other-than-zinc-blende crystal structures are also possible. 
The Hamiltonian parameters for ternary alloys, for instance, are typically calculated by using a linear combination of the parameters for the constituents. 
Therefore the alloys' parameters will be elliptic if the parameters of constituents are, because the ellipticity region is connected and convex in the space of parameters. 
Similar reasoning can be applied to the calculation of time-dependent Hamiltonian parameters with help of the Varshni formulas.
There might be some complications with analytic calculation of quadratic form's eigenvalues, for more complicated Hamiltonians with different symmetry-structure.
This is not a major issue, because for any specific material the ellipticity of $k\cdot p$ Hamiltonian can also be  verified numerically.

\section*{Acknowledgements}
The first author acknowledges the partial financial support from The Royal Society of Canada via 2017 RSC-Ukraine exchange program.
Both authors are grateful to Dr. Sunil Patil for our earlier discussions on the topic. 
The support of NSERC and the CRC Program is also acknowledged.
\bibliographystyle{apsrev4-1}
\bibliography{%
	kp,%
	sytnyk%
}

  \end{document}